\documentclass[twocolumn,aps,prc,10pt,superscriptaddress,preprintnumbers]{revtex4-1}

\usepackage[english]{babel}
\usepackage[dvipsnames,svgnames,x11names]{xcolor}
\usepackage[markup=underlined]{changes}
\usepackage{float}
\restylefloat{table}

\usepackage{graphicx}
\usepackage{amssymb}
\usepackage{amsmath}
\usepackage{epsfig}
\usepackage{bm}
\usepackage[dvipsnames]{xcolor}
\usepackage{multirow}
\usepackage{ulem}
\usepackage{xcolor}
\usepackage{comment}

\definechangesauthor[color=blue]{JS}
\definechangesauthor[color=magenta]{MN}
\definechangesauthor[color=red]{IK}
\definechangesauthor[color=green]{BT}
\definechangesauthor[color=purple]{*}

\colorlet{Changes@Color}{black!90}

\begin{document}

\title{Shear viscosity from fluctuations in linearized relativistic hydrodynamics}

\author{Jakub \v{S}t\v{e}rba}
\affiliation{Fakulta jadern\'a a fyzik\'aln\v{e} in\v{z}en\'yrsk\'a, \v{C}esk\'e vysok\'e u\v{c}en\'i technick\'e v Praze, B\v{r}ehov\'a 7, 11519 Praha 1, Czech Republic}
\affiliation{SUBATECH UMR 6457 (IMT Atlantique, Universit\'e de Nantes, IN2P3/CNRS), 4 rue Alfred Kastler, 44307 Nantes, France}
\author{Iurii Karpenko}
\affiliation{Fakulta jadern\'a a fyzik\'aln\v{e} in\v{z}en\'yrsk\'a, \v{C}esk\'e vysok\'e u\v{c}en\'i technick\'e v Praze, B\v{r}ehov\'a 7, 11519 Praha 1, Czech Republic}
\author{Marlene Nahrgang}
\affiliation{SUBATECH UMR 6457 (IMT Atlantique, Universit\'e de Nantes, IN2P3/CNRS), 4 rue Alfred Kastler, 44307 Nantes, France}
\author{Boris Tom\'a\v{s}ik}
\affiliation{Fakulta jadern\'a a fyzik\'aln\v{e} in\v{z}en\'yrsk\'a, \v{C}esk\'e vysok\'e u\v{c}en\'i technick\'e v Praze, B\v{r}ehov\'a 7, 11519 Praha 1, Czech Republic}
\affiliation{Univerzita Mateja Bela, Tajovsk\'eho 40, 974 01 Bansk\'a Bystrica, Slovakia}

\newcommand{\MN}[1]{\textcolor{magenta}{#1}}
\newcommand{\IK}[1]{\textcolor{orange}{IK: #1}}
\newcommand{\BT}[1]{\textcolor{green}{BT: #1}}
\newcommand{\JS}[1]{\textcolor{blue}{JS: #1}}

\begin{abstract}
In this work we incorporate thermal fluctuations 
into fluid-dynamic simulations with the vHLLE code, in linear approximation.
The fluctuations are built in following the fluctuation-dissipation theorem.
The effect of space discretization into cells in numerical hydrodynamics on the spatial correlation of fluctuations is studied with the help of static structure factor. 
A difference from the continuum result is pointed out and systematically investigated. 
Time evolution of the ensuing fluctuations is examined and the Green-Kubo formula is used for the extraction of shear viscosity from the fluctuations.
The extracted value 
is demonstrated to be the same as the input value of the shear viscosity for the simulations using Israel-Stewart equations.
It is also shown that numerical regulators commonly used in the codes to ensure stability of the calculation severely spoil the fluctuations and break the fluctuation-dissipation theorem.
\end{abstract}

\maketitle


\section{Introduction}
\label{s:intro}

The framework of relativistic fluid dynamics is frequently used for simulations of bulk matter evolution in relativistic heavy-ion collisions. 
The state-of-the-art models usually employ a scheme that involves dissipation and is developed up to second order in velocity gradients \cite{Israel:1976tn,Israel:1979wp,Karpenko:2013wva,Schenke:2010nt,Schenke:2010rr,Paquet:2015lta,Shen:2014vra}.
The fields, including energy density, velocity, shear stress, etc, are evolved as average values. 
The average is taken over an ensemble of systems that correspond to the same macroscopic initial conditions. 
Realistic systems, however, also exhibit stochastic thermal fluctuations around this average evolution, as dictated by the fluctuation-dissipation theorem  \cite{Landau:1980stat2}.
Their omission may be justified  if the fluctuation of the shear stress $\delta \pi$ is negligible with respect to the pressure $p$, but this is not the case in heavy-ion collisions \cite{Kapusta:2011gt}. 
A simple estimate can be based on the fluctuation-dissipation relation 
$\langle \delta \pi^2\rangle \sim 4T\eta /(\Delta t\, \Delta V)$
where $\eta$ is the shear viscosity and $\Delta t$, $\Delta V$ are the time step and cell volume, respectively, of the numerical implementation. 
Inserting relevant values gives $\delta \pi/p$ at the level of a few per cent. 

Hydrodynamic fluctuations were included in relativistic hydrodynamics in \cite{Kapusta:2011gt}, following the non-relativistic treatment of Landau and Lifshitz \cite{Landau:1980stat2}.
Since then, different approaches have been developed to describe hydrodynamic fluctuations in the context of relativistic heavy-ion collisions.
These include causal formulations of relativistic fluctuating hydrodynamics and direct stochastic or event-by-event implementations \cite{Murase:2013tma,Young:2014pka,Yan:2015lfa,Murase:2015oie,Nahrgang:2017FDFluct,Chattopadhyay:2017rgh,Singh:2018dpk,Sakai:2020pjw,Sakai:2021rug,Kuroki:2023ebq}, as well as hydro-kinetic approaches in which the fluctuations are represented by additional correlation functions following deterministic evolution equations \cite{Akamatsu:2016llw,Akamatsu:2017rdu,An:2019csj}.
More recently, a stochastic formulation based on Metropolis updates has been proposed as an alternative to the direct Langevin evolution of hydrodynamic noise \cite{Basar:2024metropolis,Bhambure:2025metropolis,Chattopadhyay:2025critical}.

The various approaches differ in how the fluctuations and their coupling to the background medium are treated.
In particular, thermal fluctuations can be evolved perturbatively on top of an averaged hydrodynamic background \cite{Young:2014pka,Chattopadhyay:2017rgh}.
In this approximation their evolution decouples from that of the averaged background.
We use this linear approximation here as well, but we also check its applicability. 
Other implementations evolve the fluctuations together with the full fluid-dynamical fields and thereby retain nonlinear interactions between fluctuations and background \cite{Murase:2013tma,Nahrgang:2017FDFluct,Singh:2018dpk}.
The hydrokinetic formulations provide yet another route by evolving correlation functions instead of individual stochastic realizations and allow the short-distance singularities associated with hydrodynamic fluctuations to be treated systematically \cite{Akamatsu:2016llw,Akamatsu:2017rdu,An:2019csj}.

It is then  a relevant question to what extent such thermal fluctuations may influence the finally observed distributions of hadrons \cite{Kapusta:2011gt,Young:2014pka,Yan:2015lfa,Sakai:2020pjw,Sakai:2021rug,Kuroki:2023ebq}? 
This could impact the commonly used paradigm, that at least the lowest orders of azimuthal anisotropies of those distributions reflect the spatial anisotropies of the energy density profile in the initial state \cite{Niemi:2012aj}.

The proper treatment of fluctuations becomes even more important if the matter evolves in the vicinity of the critical point of the QCD phase diagram, where fluctuations are expected to be enhanced \cite{Stephanov:2008qz}.
Their dynamical evolution is additionally affected by critical slowing down, which has motivated a broad range of dynamical approaches; for an overview see \cite{Bluhm:2020EMMI}.
In particular, stochastic diffusion equations have been used to describe the dynamics of the critical net-baryon density including conservation laws, nonlinear interactions, finite-size and finite-resolution effects \cite{Nahrgang:2018Diffusive,Nahrgang:2020Diffusive}.

The numerical implementation of fluctuations in relativistic viscous fluid dynamics is non-trivial and its predictive power has to be studied carefully, which is the goal of this paper.
One difficulty originates from the local form of the fluctuation-dissipation relation.
The white-noise spatial correlation is given by a $\delta$-function and, after discretisation, the amplitude of the stochastic noise scales inversely with the square root of the cell volume,
\begin{equation}
\xi\sim \frac{1}{\sqrt{\Delta V}} .
\end{equation}
Consequently, fluctuations on the scale of an individual numerical cell become stronger as the grid is refined.
This effect is already present for linear Gaussian fluctuating hydrodynamics and does not require nonlinear interactions between the fluctuations.
Nonlinear stochastic dynamics introduces additional ultraviolet cutoff dependences through the interaction of short-wavelength fluctuation modes.
Such cutoff dependences and their renormalization have been studied both for relativistic hydrodynamic fluctuations \cite{Nahrgang:2017FDFluct,Akamatsu:2016llw,Akamatsu:2017rdu,An:2019csj} and for stochastic critical dynamics \cite{Attieh:2024ModelA}.
In direct event-by-event calculations, the large short-wavelength fluctuations generated by white noise have also been regulated by introducing a finite smoothing or cutoff scale \cite{Young:2014pka,Murase:2015oie,Singh:2018dpk,Sakai:2020pjw,Sakai:2021rug}.

A related question is to what extent the numerical discretization reproduces the equilibrium fluctuations required by the fluctuation-dissipation relation.
For non-relativistic Landau--Lifshitz fluctuating hydrodynamics, the numerical properties of stochastic fluid-dynamical schemes have been studied systematically by Donev et al.\ \cite{Donev:2010}.
In particular, static structure factors provide a sensitive measure of whether the numerical evolution reproduces the equilibrium fluctuation spectrum required by the fluctuation-dissipation relation. Deviations from the continuum structure factor directly reveal modifications introduced by the discretization.
Structure factors, which are the Fourier transforms of equal-time correlation functions, have also been used in the context of heavy-ion collisions to quantify finite-resolution and finite-size effects in stochastic diffusion dynamics \cite{Nahrgang:2020Diffusive}.
We follow this strategy here and calculate static structure factors in order to study the approach of the numerical implementation towards the continuum result.

An additional complication arises from the stability requirements of viscous fluid-dynamical algorithms.
In practical implementations, the viscous correction $\pi^{\mu\nu}$ is commonly limited when it becomes too large compared with the ideal part of the energy-momentum tensor.
Such a prescription is useful for stabilizing the fluid-dynamical evolution, but becomes problematic when large values of $\pi^{\mu\nu}$ originate from thermal fluctuations.
Since the typical magnitude of the noise increases when $\Delta V$ is reduced, the regulator becomes active more frequently as the numerical grid is refined.
The resulting modification of the fluctuating viscous stress is not accompanied by the corresponding modification of the dissipation and therefore spoils the fluctuation-dissipation relation.
Since such regulators are usually required in realistic viscous-hydrodynamic simulations, it is essential to establish whether they remain compatible with the fluctuation-dissipation relation.

In this article we construct and study a numerical implementation of a relativistic viscous fluid-dynamical model with thermal fluctuations.
In order for the results to be analytically traceable, the fluctuations are treated as linear perturbations on top of the averaged background.
Furthermore, we limit our benchmarks to the evolution of fluctuations on top of a homogeneous background medium, in a static box with periodic boundary conditions.
This setup allows us to disentangle properties of the stochastic numerical implementation from effects caused by the space-time evolution of the background.

The Green--Kubo formula \cite{Kubo:1966fyg,Hosoya:1983id} has been used earlier in transport simulations \cite{Demir:2008tr,Wesp:2011yy,Rose:2020lfc,Hammelmann:2023fqw} to obtain the shear viscosity from the fluctuations of the shear stress tensor.
Here, we use the Green--Kubo relation as an independent consistency test of the fluctuation-dissipation relation in fluctuating relativistic fluid dynamics.
We demonstrate that the shear viscosity used as input to the fluid-dynamical evolution can be recovered from the equilibrium shear-stress correlation function.
We also investigate the breakdown of this relation when the viscous regulator becomes active, which is a practically relevant case in event-by-event stochastic simulations.

The paper is organised as follows. 
The model is introduced in the next Section. 
We explain how linear equations for small perturbations over the background fields (energy density, velocity, shear stress) are formulated  (Sec.~\ref{s:linpert}) and how stochastic noise is implemented (Sec.~\ref{s:noise}).
Separate Subsection~\ref{s:nnum} is devoted to specific artifacts that stem from the discretisation of the equations, since it has important impact on the results. 
We specify how our simulations are performed in Sec.~\ref{s:sims}.
In Section~\ref{s:sf} we calculate the static structure factors of the fluctuating fields, both analytically where possible (Sec.~\ref{ss:anal1} and Appendix \ref{a:anal-SF}), and numerically from our simulations (Sec.~\ref{s:numss}). 
Our main results are presented in Sec.~\ref{s:visc} where we measure viscosity from the Green-Kubo formula and compare it with the value that was put into the simulations. 
Conclusions are summarised in Section~\ref{s:conc}. 
Technical details are deferred to the appendices.


\section{Relativistic fluid dynamics with thermal fluctuations}
\label{s:model}

The equations governing the hydrodynamic evolution are the energy-momentum conservation
\begin{equation}
    \partial_\nu T^{\mu\nu}=0
\end{equation}
together with the relaxation Israel-Stewart equations \cite{Israel:1976tn,Israel:1979wp}
\begin{subequations}
\label{e:IS_eq}
\begin{eqnarray}
    \langle u^{\gamma}\partial_\gamma\pi^{\mu\nu}\rangle_\perp&=&-\frac{\pi^{\mu\nu}-\pi^{\mu\nu}_{NS}}{\tau_\pi}-\frac{\delta_{\pi\pi}}{\tau_\pi}\pi^{\mu\nu}\partial_\alpha u^\alpha
\nonumber \\ & & 
+\frac{\phi_7}{\tau_\pi}\pi_\alpha^{\langle\mu}\pi^{\nu\rangle_\perp\alpha}
    -\frac{\tau_{\pi\pi}}{\tau_\pi}\pi_\alpha^{\langle\mu}\sigma^{\nu\rangle_\perp\alpha}
\nonumber \\ & & 
    +\frac{\lambda_{\pi\Pi}}{\tau_\pi}\Pi\sigma^{\mu\nu}
    \\
u^{\gamma}\partial_\gamma\Pi&=&-\frac{\Pi-\Pi_{NS}}{\tau_\Pi}-\frac{\delta_{\Pi\Pi}}{\tau_\Pi}\Pi\partial_\alpha u^\alpha
\nonumber \\ & & 
+\frac{\lambda_{\Pi\pi}}{\tau_\Pi}\pi^{\mu\nu}\sigma_{\mu\nu}\, .
\end{eqnarray}
\end{subequations}
Here, the shear stress and bulk pressure, $\pi^{\mu\nu}$ and $\Pi$, are 
independent dynamical variables obeying the equations of motion above. 
They relax on characteristic time scales $\tau_\pi$ and $\tau_\Pi$ towards their Navier-Stokes values.
The symbol $\langle \dots \rangle_{\perp}$ stands for the symmetric and traceless projection perpendicular to four-velocity
\begin{multline}
\label{e:operator}
\langle A^{\mu\nu}\rangle_{\perp} = A^{\langle \mu\nu\rangle_\perp} 
\\
=\bigg[ \frac{1}{2}(\Delta^{\mu\alpha}\Delta^{\nu\beta} + \Delta^{\mu\beta} \Delta^{\nu\alpha}) - \frac{1}{3}\Delta^{\mu\nu}\Delta^{\alpha\beta} \bigg]A_{\alpha\beta}\, ,
\end{multline}
where the projector $\Delta^{\mu\nu}$ is defined as:
\[
\Delta^{\mu\nu} = g^{\mu\nu}-u^\mu u^\nu\,  .
\]
The Navier-Stokes values of the shear stress and bulk pressure are calculated as
\begin{eqnarray}
\pi^{\mu\nu}_{NS} & = & 2\eta \sigma^{\mu\nu} 
\nonumber \\
& = & \eta \left ( \Delta^{\mu\lambda} \partial_\lambda u^\nu + 
\Delta^{\nu\lambda} \partial_\lambda u^\mu \right ) - \frac{2}{3} \eta \Delta^{\mu\nu}\partial_\lambda u^\lambda \\
\Pi_{NS} & = & - \zeta \partial_\mu u^\mu \, , 
\end{eqnarray}
which also defines the strain rate tensor $\sigma^{\mu\nu}$. 
The shear and bulk viscosity are denoted by $\eta$ and $\zeta$, respectively. 

Relaxation time $\tau_\pi$ is assumed to be proportional to the viscosity
\begin{equation}
\label{e:tau-visc}
\tau_\pi  =  5 \frac{\eta}{sT}\, .
\end{equation}
In our simulations we set the bulk viscosity to zero, consequently $\tau_\Pi=0$.

Other symbols that appear in Eqs.~(\ref{e:IS_eq}) are explained in Appendix~\ref{a:symbols}.

\subsection{Linearized perturbations}
\label{s:linpert}

 The evolution equations are nonlinear, generally.  
 To simplify them, we will introduce thermal fluctuations as small perturbations and perform  linear expansion around a background.
 
 The total energy-momentum tensor is decomposed into the background part and the fluctuating contribution, assumed to be small compared to the background:
 \[ T^{\mu\nu}=T_0^{\mu\nu}+\delta T^{\mu\nu}\, . \] 
 The energy-momentum conservation then reads
 \begin{equation}
  \partial_\nu T^{\mu\nu} = 
  \partial_\nu (T_0^{\mu\nu}+\delta T^{\mu\nu})=0
 \end{equation}
 The background is given by the ensemble average of the energy momentum-tensor
\begin{equation}
\langle T^{\mu\nu}\rangle=\langle T_0^{\mu\nu}+\delta T^{\mu\nu}\rangle=T_0^{\mu\nu}\, ,
\end{equation}
thus the fluctuations satisfy
\begin{equation}
    \langle\delta T^{\mu\nu}\rangle=0.
\end{equation}
The energy-momentum conservation is then valid also for the ensemble average
\begin{equation}
\partial_\nu \langle T^{\mu\nu}\rangle = 
\partial_\nu T_0^{\mu\nu} = 0\, ,
\label{e:EC0}
\end{equation}
and the fluctuating part fulfills separately
\begin{equation}
\partial_\nu  \delta T^{\mu\nu}  =  0\, .
\label{e:ECfluct}
\end{equation}
Linear fluctuations only contribute to Eq.~(\ref{e:ECfluct}) and not to the evolution of the averaged background in Eq.~(\ref{e:EC0}). 
The background, on the other hand, will be shown to affect the fluctuations. 

To formulate the energy–momentum tensor explicitly, intensive thermodynamic variables, flow velocity as well as the shear stress and bulk pressure are decomposed into background and fluctuation
\begin{equation}
\begin{split}
        \varepsilon&=\varepsilon_0+\delta\varepsilon \\
        p&=p_0+\delta p \\
        u^\mu&=u_0^\mu+\delta u^\mu\\
        \pi^{\mu\nu}&=\pi_0^{\mu\nu}+\delta\pi^{\mu\nu}\\
        \Pi&=\Pi_0+\delta\Pi.
\label{eq:lin}
\end{split}
\end{equation}
In this approximation, we then have $\delta u_\mu u^\mu = \delta u_\mu u^\mu_0 = 0$. 
For the special case of zero background velocity ($u_0^i = 0$), 
the zeroth component of the  four-velocity fluctuation vanishes. 
Using Eq.~(\ref{eq:lin}) and retaining terms up to linear order, the two parts of the energy–momentum tensor become
\begin{eqnarray}
        T_0^{\mu\nu}&= &(\varepsilon_0+p_0)u_0^\mu u_0^\nu-p_0g^{\mu\nu}-\Pi_0\Delta_0^{\mu\nu}+\pi_0^{\mu\nu}
        \nonumber \\ \nonumber 
        \delta T^{\mu\nu}&= &(\varepsilon_0+p_0)(u_0^\mu\delta u^\nu+u_0^\nu\delta u^\mu)+(\delta\varepsilon+\delta p)u_0^\mu u_0^\nu \\
        && {} -\delta pg^{\mu\nu} -\delta\Pi\Delta_0^{\mu\nu} 
        + \Pi_0 (u_0^\mu \delta u^\nu + u_0^\nu \delta u^\mu ) 
        \nonumber \\ 
        && {} + \delta\pi^{\mu\nu}
\label{e:delTmunu}
\end{eqnarray}
where 
\[
\Delta_0^{\mu\nu}=g^{\mu\nu}-u_0^\mu u_0^\nu\\
\]
is a projector orthogonal to the background four-velocity  $u_0^\mu$.

The Israel-Stewart equations (\ref{e:IS_eq}) are linearized in an analogous manner using Eq.~(\ref{eq:lin}). 
In addition to this, the transport coefficients may depend on temperature and hence we also decompose
\begin{equation}
\begin{split}
\eta & = \eta_0 + \delta \eta \\
\zeta & = \zeta_0 + \delta \zeta \\
\tau_\pi & = \tau_{\pi,0} + \delta \tau_\pi \\
\tau_\Pi & = \tau_{\Pi,0} + \delta \tau_\Pi \\
\phi_7 & = \phi_{7,0} + \delta \phi_7 \, .
\end{split}
\end{equation}
The zeroth order equations look formally like Eqs.~(\ref{e:IS_eq}), just with background quantities 
\begin{subequations}
\label{e:IS0}
\begin{eqnarray}
\langle u^{\gamma}_0\partial_\gamma\pi^{\mu\nu}_0\rangle_{0\perp}&=&
-\frac{\pi_0^{\mu\nu}-\pi^{\mu\nu}_{0,NS}}{\tau_{\pi,0}}
-\frac{\delta_{\pi\pi}}{\tau_{\pi}}
\pi^{\mu\nu}_0\partial_\alpha u^\alpha_0
\nonumber \\ & & 
{}+\frac{\phi_{7,0}}{\tau_{\pi,0}}\pi_{0,\alpha}^{\phantom{0\alpha}\langle\mu}
\pi_0^{\nu\rangle_{0\perp}\alpha}
-\frac{\tau_{\pi\pi}}{\tau_{\pi}}
\pi_{0,\alpha}^{\phantom{0\alpha}\langle\mu}\sigma^{\nu\rangle_{0\perp}\alpha}
\nonumber \\ & & 
{}    +\frac{\lambda_{\pi\Pi}}{\tau_{\pi}}\Pi_0\sigma_0^{\mu\nu}
    \\
u^{\gamma}_0\partial_\gamma\Pi_0&= & 
-\frac{\Pi_0-\Pi_{0,NS}}{\tau_{\Pi,0}}
-\frac{\delta_{\Pi\Pi}}{\tau_{\Pi}}
\Pi_0\partial_\alpha u_0^\alpha
\nonumber \\ & & 
{} +\frac{\lambda_{\Pi\pi}}{\tau_{\Pi}}
\pi^{\mu\nu}_0\sigma_{0,\mu\nu}\, .
\end{eqnarray}
\end{subequations}
The symmetric and traceless projection is now oriented perpendicularly to the background velocity
\begin{multline}
\label{e:operator0}
\langle A^{\mu\nu}\rangle_{0\perp} = A^{\langle \mu\nu \rangle_{0\perp}} 
\\
=\bigg[ \frac{1}{2}(\Delta_0^{\mu\alpha}\Delta_0^{\nu\beta} + \Delta_0^{\mu\beta} \Delta_0^{\nu\alpha}) - \frac{1}{3}\Delta_0^{\mu\nu}\Delta_0^{\alpha\beta} \bigg]A_{\alpha\beta}\, ,
\end{multline}

Symbols and quantities not explained here can be found in Appendix~\ref{a:symbols}.

The evolution equations for the fluctuating shear stress and bulk pressure are
\begin{widetext}
\begin{subequations}
\label{e:IS_fluc}
\begin{eqnarray}
    u_0^\lambda\partial_\lambda\delta\pi^{\mu\nu}&= & 
    -\frac{\delta\pi^{\mu\nu}-\delta\pi^{\mu\nu}_{NS}}{\tau_{\pi,0}}
    +\frac{\pi_0^{\mu\nu}-\pi^{\mu\nu}_{0,NS}}{\tau_{\pi,0}^2}\delta\tau_\pi
    -\delta u^\gamma\partial_\gamma\pi_0^{\mu\nu}
    -[\pi_0^{\mu\beta}\delta u^\nu+\delta\pi^{\mu\beta}u_0^\nu+\pi_0^{\beta\nu}\delta u^\mu+\delta\pi^{\beta\nu}u_0^\mu]u_0^\gamma\partial_\gamma u_{0,\beta}
\nonumber \\ &&    
    {}-[\pi_0^{\mu\beta}u_0^\nu+\pi_0^{\beta\nu}u_0^\mu](u_0^\gamma\partial_\gamma\delta u_\beta+\delta u^\gamma\partial_\gamma u_{0,\beta})
    -\frac{\delta_{\pi\pi}}{\tau_{\pi}}(\pi_0^{\mu\nu}\partial_\gamma\delta u^\gamma+\delta\pi^{\mu\nu}\partial_\gamma u_0^\gamma)
\nonumber    \\ &&
    {}+\frac{\phi_{7,0}}{\tau_{\pi,0}}\bigg[ \frac{1}{2} (\delta\pi_\alpha^\mu \pi_0^{\nu\alpha}+\pi_{0,\alpha}^\mu\delta\pi^{\nu\alpha}+\delta\pi_\alpha^\nu\pi_0^{\mu\alpha}+\pi_{0,\alpha}^\nu\delta\pi^{\mu\alpha}) -\frac{2}{3}\Delta_0^{\mu\nu}\delta\pi_{\alpha\beta}\pi_0^{\alpha\beta}
+\frac{1}{3}(\delta u^\mu u_0^\nu+u_0^\mu\delta u^\nu)\pi_{0,\alpha\beta}\pi_0^{\alpha\beta}\bigg]
\nonumber  \\ && 
+\bigg(\frac{\delta\phi_{7}}{\tau_{\pi,0}}-\frac{\phi_{7,0}}{\tau_{\pi,0}^2}\delta\tau_\pi\bigg)\pi_{0,\alpha}^{\langle\mu}\pi_0^{\nu\rangle\alpha}
-\frac{\tau_{\pi\pi}}{\tau_{\pi,0}}
    \bigg[ \frac{1}{2} (\delta\pi_\alpha^\mu \sigma_0^{\nu\alpha}+\pi_{0,\alpha}^\mu\delta\sigma^{\nu\alpha}+\delta\pi_\alpha^\nu\sigma_0^{\mu\alpha}+\pi_{0,\alpha}^\nu\delta\sigma^{\mu\alpha})
    \nonumber \\ && \phantom{10} {}
    -\frac{1}{3}\Delta_0^{\mu\nu}(\delta\pi_{\alpha\beta}\sigma_0^{\alpha\beta}
    +\pi_{0,\alpha\beta}\delta\sigma^{\alpha\beta})
    +\frac{1}{3}(\delta u^\mu u_0^\nu+u_0^\mu\delta u^\nu)\pi_{0,\alpha\beta}\sigma_0^{\alpha\beta}\bigg]
\nonumber \\ &&
    +\frac{\lambda_{\pi\Pi}}{\tau_{\pi}}(\Pi_0\delta\sigma^{\mu\nu}+\delta\Pi\sigma_0^{\mu\nu})\, ,
    \\
u_0^\gamma\partial_\gamma\delta\Pi&=&
-\frac{\delta\Pi-\delta\Pi_{NS}}{\tau_{\Pi,0}}-\frac{\Pi_0-\Pi_{0,NS}}{\tau_{\Pi,0}^2}\delta\tau_\Pi-
\frac{\delta_{\Pi\Pi}}{\tau_{\Pi}}(\Pi_0\partial_\alpha \delta u^\alpha+\delta\Pi\partial_\alpha u_0^\alpha)
+\frac{\lambda_{\Pi\pi}}{\tau_{\Pi}}(\pi_0^{\mu\nu}\delta\sigma_{\mu\nu}+ \delta\pi_{\mu\nu}\sigma_0^{\mu\nu})\, .
\end{eqnarray}
\end{subequations}
\end{widetext}
The expressions introduced here are also defined in Appendix~\ref{a:symbols}. 
Note that the ratios $\delta_{\pi\pi}/\tau_\pi$, $\tau_{\pi\pi}/\tau_\pi$, and $\tau_{\pi\Pi}/\tau_\pi$ are pure numbers, thus their variations vanish, see Eq.~(\ref{e:coefs}).

Equations (\ref{e:delTmunu}), (\ref{e:IS0}), and (\ref{e:IS_fluc}) demonstrate the earlier statement that background evolution directly affects the fluctuations, while the fluctuations do not produce any back-reaction on the background at this order.
The evolution of the background is independent of fluctuations in this approximation.  

The evolution equations simplify considerably in a static background, where the fields are homogeneous and $[u_0^\mu] = (1,0,0,0)$. The background shear stress and bulk pressure then vanish, and the Israel-Stewart equations become
\begin{subequations}
\label{e:ISflucev}
\begin{eqnarray}
\partial_t \delta\pi^{\mu\nu} & = & 
- \frac{\delta\pi^{\mu\nu} - \delta\pi^{\mu\nu}_{NS}}{\tau_{\pi,0}}
\\
\partial_t \delta\Pi & = & 
- \frac{\delta\Pi - \delta\Pi_{NS}}{\tau_{\Pi,0}}
\end{eqnarray}
\end{subequations}
with the Navier-Stokes values
\begin{eqnarray}
\delta\pi^{\mu\nu}_{NS} & =  & \eta_0 
\left [ \Delta_0^{\mu\lambda} \partial_\lambda \delta u^\nu  
+ \Delta_0^{\nu\lambda} \partial_\lambda \delta u^\mu 
- \frac{2}{3} \Delta_0^{\mu\nu} \partial_\lambda \delta u^\lambda
\right ]
\nonumber \\ && \\
\delta \Pi_{NS} & = & - \zeta_0 \partial_\lambda \delta u^\lambda\, .
\end{eqnarray}


\subsection{Stochastic noise}
\label{s:noise}

Thermal fluctuations will be introduced through a stochastic noise term $\Xi^{\mu\nu}$, which is added to the dissipative part of the energy-momentum tensor $T^{\mu\nu}_{visc}$  
\cite{Young:2013fka}
\begin{equation}
    T^{\mu\nu}=T^{\mu\nu}_{id}+T^{\mu\nu}_{visc} + \Xi^{\mu\nu} = T^{\mu\nu}_{id}+T^{\prime\mu\nu}_{visc}\,  .
\end{equation}
In the second equality the noise is absorbed into a redefinition of the dissipative part.
Following \cite{Young:2013fka} the noise term obeys relaxation equation
\begin{equation}
\partial_t \Xi^{\mu\nu} = -\frac{\Xi^{\mu\nu} - \xi^{\mu\nu}}{\tau_\xi}\, ,
\end{equation}
where the white noise $\xi^{\mu\nu}$ and the relaxation time $\tau_\xi$ will be specified below. 
Analogously to the viscous corrections, the noise terms can be decomposed into transverse traceless and scalar components, $\Xi_\pi^{\mu\nu}$ and $\Xi_\Pi$, and used to redefine the shear stress and bulk pressure
\begin{subequations}
\label{e:primed}
\begin{eqnarray}
    \pi^{\prime\mu\nu}& =& \pi_0^{\mu\nu} + \delta{\pi'}^{\mu\nu}
    =\pi_0^{\mu\nu}+ \delta{\pi}^{\mu\nu}+\Xi_\pi^{\mu\nu}
    \\
    \Pi^\prime&=&\Pi_0 + \delta\Pi' = \Pi_0 + \delta\Pi + 
    \Xi_\Pi
\end{eqnarray}
\end{subequations}
Accordingly, the white noise is also decomposed
\begin{equation}
\xi^{\mu\nu} = \xi_\pi^{\mu\nu} - \Delta_0^{\mu\nu} \xi_\Pi\, .
\end{equation}
the $\xi^{\mu\nu}_\pi$ being the traceless part and $\xi_\Pi$ the non-traceless part of the white noise.
Following the procedure from \cite{Young:2013fka}, the fluctuation-dissipation theorem determines the correlator of the white noise traceless component, 
\begin{multline}
        \langle \xi_\pi^{\mu\nu}(x)\xi_\pi^{\alpha\beta}(x^\prime) \rangle = 2 T_0\eta_0 
        \\ \times
        \bigg[ \Delta_0^{\mu\alpha}\Delta_0^{\nu\beta}+\Delta_0^{\mu\beta}\Delta_0^{\nu\alpha}
        -\frac{2}{3} \Delta_0^{\mu\nu}\Delta_0^{\alpha\beta} \bigg] 
        \delta^4(x-x^\prime)
    \label{e:xicor}
\end{multline}
as well as the non-traceless component 
\begin{equation}
    \langle\xi_\Pi(x)\xi_\Pi(x')\rangle=2\zeta_0 T_0\delta^4(x-x^\prime)\, .
\label{e:xiPcor}
\end{equation}
Recall that  angular brackets $\langle\dots\rangle$ denote ensemble averaging.

The Israel–Stewart equations for the fluctuating part (\ref{e:IS_fluc}) then change by adding the noise terms to the first ratios on the r.h.s.
\begin{equation}
\label{e:IS_stoch}
\begin{split}
    u_0^\lambda\partial_\lambda\delta\pi^{\prime\mu\nu}&=
    -\frac{\delta\pi^{\prime\mu\nu}-\delta\pi^{\mu\nu}_{NS}-\xi_\pi^{\mu\nu}}{\tau_{\pi,0}}-\dots\\
    u_0^\gamma\partial_\gamma\delta\Pi^\prime&=-\frac{\delta\Pi^\prime-\delta\Pi_{NS}-\xi_\Pi}{\tau_{\Pi,0}}-\dots ,
\end{split}
\end{equation}
where the ellipses mean the remaining terms from the equations~(\ref{e:IS_fluc}).
Note that for the case with \emph{static background}, the r.h.s. includes only these terms that are listed here, so the modification of Eqs.~(\ref{e:ISflucev}) reads
\begin{subequations}
\begin{eqnarray}
\label{e:v0delpi}
\partial_t \delta\pi^{\prime\mu\nu} & = & 
- \frac{\delta\pi^{\prime\mu\nu} - \delta\pi^{\mu\nu}_{NS}-\xi_\pi^{\mu\nu}}{\tau_{\pi,0}}
\\
\partial_t \delta\Pi^\prime & = & 
- \frac{\delta\Pi^\prime - \delta\Pi_{NS}-\xi_\Pi}{\tau_{\Pi,0}}
\end{eqnarray}
\end{subequations}


\subsection{Numerical implementation}
\label{s:nnum}

The procedure used to generate noise fields satisfying the correlator structure in Eqs.~(\ref{e:xicor}) and (\ref{e:xiPcor}) follows the method proposed in \cite{Young:2013fka}.
For the actual implementation in numerical hydrodynamics, 
noise contributions in a given cell at a given time step are generated from Gaussian distribution with the width given by
\begin{eqnarray}
        \langle \xi_\pi^{\mu\nu}(x)\xi_\pi^{\alpha\beta}(x) \rangle & = &
        \frac{2T_0\eta_0}{\Delta x^3\, \Delta t} 
        \bigg[ \Delta_0^{\mu\alpha}\Delta_0^{\nu\beta}+\Delta_0^{\mu\beta}\Delta_0^{\nu\alpha}
        \nonumber \\ & & 
        -\frac{2}{3} \Delta_0^{\mu\nu}\Delta_0^{\alpha\beta} \bigg]
\label{e:xipiwidth}\\
    \langle\xi^2_\Pi(x)\rangle &= &\frac{2\zeta_0 T_0}{\Delta x^3\, \Delta t}\, ,
\label{e:xiPiwidth}
\end{eqnarray}
where $\Delta x$ is the size of the cell and $\Delta t$ the time step. 
Note that in order to compare to analytic results we will also run one-dimensional simulations where the normalisation in the denominators is changed accordingly to $\Delta x\Delta t$.

The Israel-Stewart scheme of relativistic fluid dynamics assumes that the shear stress $\pi^{\mu\nu}$ and bulk pressure $\Pi$ are small perturbations on the ideal energy-momentum tensor $T_0^{\mu\nu}$. 
The scheme breaks down when the components of these terms become comparable with the pressure. 

Each numerical implementation of relativistic fluid dynamics has to deal with such a situation, because the fluxes of energy and momentum density between the neighboring cells could even cause non-physical values of the fields, e.g., negative energy density in a cell. 

This can easily happen if there are white noise contributions to the shear stress and bulk pressure according to Eq.(\ref{e:primed}). 
As follows from Eqs.~(\ref{e:xipiwidth}) and (\ref{e:xiPiwidth}), the amplitude of the noise is inversely proportional to $\Delta x^{\frac{3}{2}}$ and so this happens more often for discretisations with smaller cells. 

Our simulation is based on the vHLLE code \cite{Karpenko:2013wva}. 
We set the bulk viscosity to zero, hence there is no  $\xi_\Pi$.
To avoid unphysically high shear stress, the highest value of its components is compared with the highest value among energy-momentum tensor components.
If the former is higher, then $\pi_{\mu\nu}$ is set to 1/10 of its original value.
Such a \emph{limiter} (or \emph{viscous regulator}\!)---like any other limiter that could be implemented \cite{Dusling:2007gi}---spoils the compatibility of the noise generation with the fluctuation-dissipation theorem. 
Nevertheless, it must be included in order to avoid unstable solutions, and any hydrodynamic code must include it in some form. 
Furthermore, our simulation also checks that the total energy density $(\varepsilon_0 + \delta \varepsilon)$ remains positive and prevents $\delta\varepsilon$ from becoming larger than the background.
This is consistent with the assumption that in linear hydrodynamics fluctuations are always smaller than the background fields.
Below, we will also study its impact on the simulated fluctuations. 

In this work we will examine fluctuations over a static background.
In such a case, thermal fluctuations do not compete with any other source of inhomogeneities and  we can avoid  invoking the limiter by a simple scaling trick: each time the noise contribution $\xi_\pi^{\mu\nu}$ is generated, it is multiplied by a factor $\alpha \ll 1$, in order to decrease it and keep $\pi^{\mu\nu}$ small, such that there is no need to invoke the limiter.
Then, when correlation functions of the type $\langle \pi^{\mu\nu} \pi^{\kappa\lambda}\rangle $ are evaluated, they are multiplied back by a factor $\alpha^{-2}$ in order to restore their actual size. 
Since for the static background,  thermal noise is the only source of shear stress and bulk pressure and the relaxation equations (\ref{e:ISflucev}) are linear in fluctuations, this trick can be used. 
In case of dynamically evolving background, e.g.\ an expanding fireball, the limiter is the only way to avoid the non-physical values of the fields. 


\section{Simulations}
\label{s:sims}

Our calculations and simulations in this work are done with homogeneous static background with periodic boundary conditions. 
We choose the equation of state of an ultra-relativistic gas 
\begin{equation}
p = \frac{\varepsilon}{3}\, .
\end{equation}
The energy density is related to temperature
\begin{equation}
\varepsilon = \frac{\pi^2}{30}g T^4\, ,
\end{equation}
with $g = 2(N_c^2-1)+\frac{7}{8} 4 N_c N_f$ and $N_c = 3$, $N_f = 2.5$. 

Note that we do not evolve the baryon density. 
The background energy density is set to 1~GeV/fm$^3$. 
Shear viscosity is always proportional to the entropy density
\begin{equation}
    \frac{\eta}{s} = \mbox{const}\, ,
\end{equation}
and several values of this ratio will be probed. 
The entropy density is obtained as 
\begin{equation}
    s = \frac{p+\varepsilon}{T}\, .
\end{equation}

In numerical simulations the system is discretised into a number of cells in each direction. 
The evolution proceeds in time steps. 

Any simulation starts with the fluctuating fields set to 0. 
The shear stress is set to the background Navier-Stokes value, but for constant background this vanishes, as well. 
Since the fluctuations are generated by the white noise, we always run the evolution for some time in order to equilibrate the fields, and perform our measurements only after that. 


\section{Structure factors}
\label{s:sf}

In actual numerical simulation, space 
is discretised and this influences the correlations of the thermal noise, which is defined in continuum via Eqs.~(\ref{e:xicor}) and (\ref{e:xiPcor}).
It is important for applications to understand to what extent the simulated fluctuations indeed correspond to those realised by Nature. 
Note that the task will be even more important when critical long-distance fluctuations are examined close to the critical point of the phase diagramme, but this is beyond our scope in the present study. 
Here we only deal with the thermal non-critical baseline fluctuations. 

The customary tool to study the structure of a field is the \emph{structure factor}.
The field is a quantity that is evolved in the problem: this can be in our case the energy density, velocity, or the shear stress.
The static structure factor provides a convenient way to characterize the equilibrium fluctuations in Fourier space. It allows us to test whether the numerical implementation reproduces the fluctuation spectrum required by the fluctuation-dissipation relation and how this spectrum approaches the continuum prediction as the grid is refined.
To define the structure factor, we introduce the Fourier transform of a field $x(\vec r,t)$
\begin{equation}
\label{e:FT}
    \hat{x}(\vec{k},t)=\mathcal{F}[x(\vec{r},t)]=\frac{1}{V}\int_{\vec{r}\in|V|}e^{-i\vec{k}\cdot\vec{r}}x(\vec{r},t)\textrm{d}\vec{r}
\end{equation}
where the integration runs over the whole examined volume $V$. 
The \emph{dynamic structure factor} is defined as the two–time correlation function
\begin{equation}
    S(\vec{k},\tau)=V\langle \hat{x}(\vec{k},t_0)\hat{x}^\ast(\vec{k},t_0+\tau)\rangle
\end{equation}
where the averaging also runs over the initial  time $t_0$. 
The \emph{static structure factor} corresponds then to the equal-time correlator
\begin{equation}
    S(\vec{k})=V\langle \hat{x}(\vec{k},t)\hat{x}^\ast(\vec{k},t)\rangle\, .
\end{equation}

\subsection{Analytic 1D static structure factor}
\label{ss:anal1}

In order to have a reference, we first calculate the static structure factor in continuum. 
For simplicity, we perform the calculation in one dimension.
This is done in such a way that all fields are allowed to vary only in one spatial direction, hence all derivatives in the other two directions drop out.
What remains is effectively one-dimensional evolution. 
The evolved fields are the fluctuations of energy density, velocity, and longitudinal component of the shear stress, from which their Fourier transforms are determined
\begin{equation}
    [\hat{x}(k,t)] = \left (
    \begin{array}{c}
    \widehat{\delta \varepsilon}(k,t) \\ \widehat{\delta u}(k,t) \\ 
    \widehat{\delta \pi}(k,t)
    \end{array}
    \right )\, ,
\end{equation}
where $u = u^x$ and $\delta\pi = \delta\pi^{\prime xx}$.
Then, the structure factor can be summarised as a matrix
\begin{equation}
    S_{ij}(k) = V \langle \hat{x}^\dagger_i(k)\hat{x}_j(k)\rangle\, .
\end{equation}
Calculation follows the procedure set out in \cite{Donev:2010} and is outlined in Appendix~\ref{a:anal-sf-cont}. 
The result is
\begin{equation}
\label{e:anal_SF}
S_{ij}(k) = \left (
\begin{array}{ccc} 
\frac{(\varepsilon_0+p_0)T_0}{c_s^2} & 0 & 0 \\
0 & \frac{T_0}{\varepsilon_0 + p_0} & 0 \\
0 & 0 & \frac{4}{3} \frac{\eta_0 T_0}{\tau_{\pi,0}}
\end{array}
    \right )
\end{equation}
Note that since there is no length scale introduced in this problem, the structure factor is a constant and does not depend on $\vec k$, reflecting the spatially white character of the noise field.

In the numerical simulation, the finite grid spacing $\Delta x$ introduces a shortest resolved length scale, or equivalently an ultraviolet cutoff in wave number. 
The structure factor therefore acquires a dependence on $\vec{k}$, in particular at wavelengths approaching the grid scale.
Detailed analytical calculation for the structure factor of a fluid discretised in cells  is again deferred to the Appendix~\ref{a:sol1d}.
The procedure ends up with the matrix equation (\ref{e:e4numSF}) from which the structure factor is obtained numerically.
Note that the calculation assumes piecewise constant distributions within the cells.
This does not exactly correspond to the default implementation in vHLLE, where the fields are taken as changing linearly over the volume of a single cell, in order to improve the numerical accuracy. 
We will therefore run some of the hydrodynamical simulations with piecewise constant distributions.

The resulting curves from the semi-analytical calculation will be compared with the results of the simulations. 


\subsection{Numerical simulations of the static structure factor}
\label{s:numss}

\begin{figure}
    \centering
    \includegraphics[width=\linewidth]{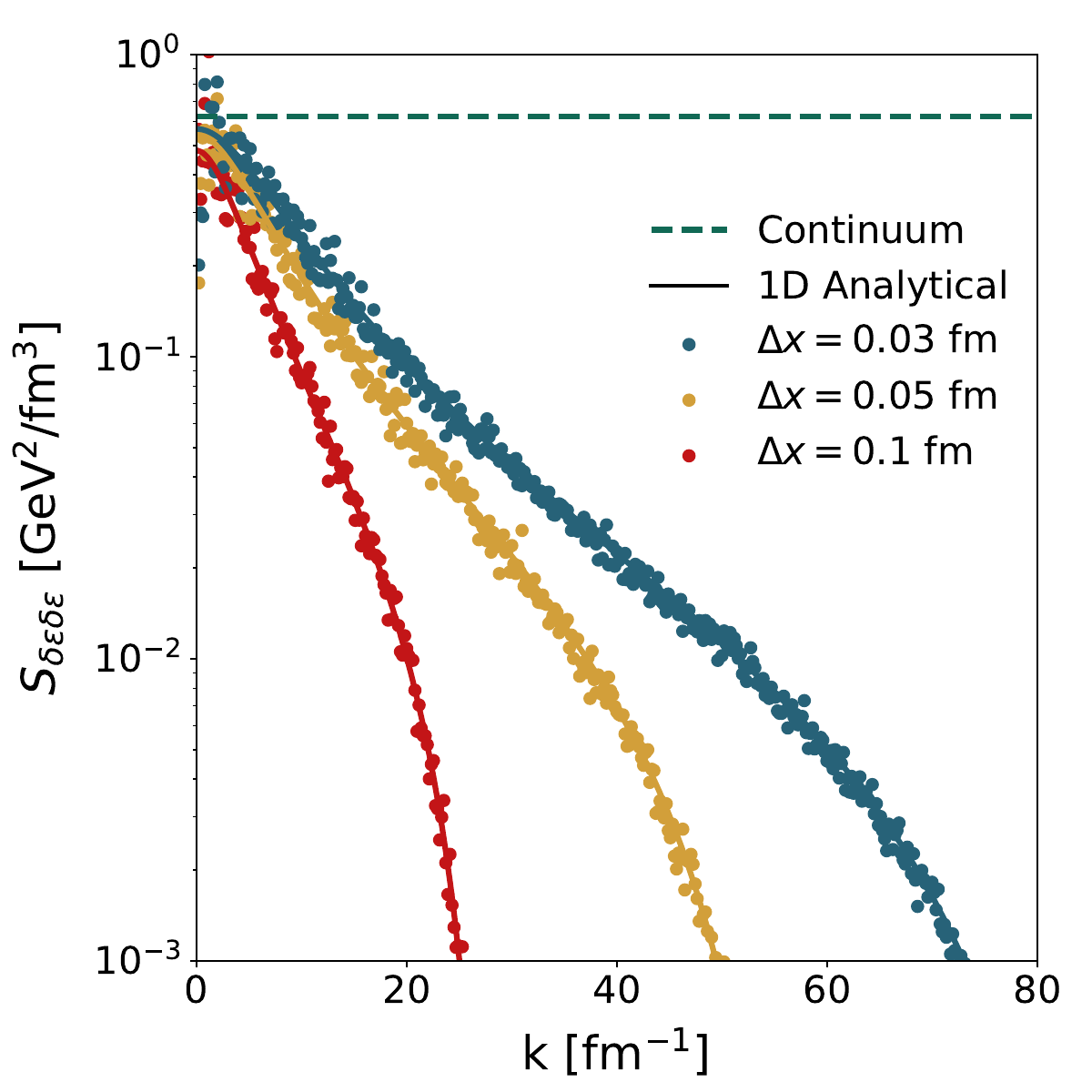}
    \caption{One-dimensional static Israel-Stewart structure factor for energy density fluctuations. The analytical continuum result according to Eq.~(\ref{e:anal_SF}) plotted with green dashed line. Data points show results from simulations with piecewise constant energy density profile within cells, $\eta/s = 0.12$, and time step $\Delta t = 0.01$~fm/$c$, for three different cell sizes. Solid lines show the semi-analytical results for the discretised numerical scheme.}
    \label{fig:1D_flat}
\end{figure}

In Fig.~\ref{fig:1D_flat}
we present the static structure factor for $\delta\varepsilon \delta\varepsilon$ correlation in one dimension.
The ratio of shear viscosity to entropy density was set to $\eta/s = 0.12$.
The static medium with periodic boundary conditions was let to evolve and the Fourier-transformed correlations were measured as a function of $k$.
Piecewise-constant distributions within the cells were assumed in order to be consistent with the semi-analytic results. 
Thermal fluctuations are scaled down by factor $\alpha=0.01$ and the structure factors are scaled up by the factor $\alpha^{-2}$, in order to prevent unphysical energy density in simulations, as explained in Section~\ref{s:nnum}. 
Numerical results are plotted with points which are scattered because of uncertainty due to limited statistics. 
The points agree very well over several orders of magnitude (even beyond the range shown in the figure) with the semi-analytic results derived for the finite cell size.
Three sizes of cells are probed. 
Generally, the structure factor decreases towards 0 as $k$ is increased. 
The larger the cell size is, the faster it decreases. 
This is expected since small cell sizes should be closer to the continuum limit.
Note that for $k=0$ the structure factor always has to vanish. 
This is dictated by $\widehat{\delta\varepsilon}(k=0) = 0$, due to energy conservation. The deviation of the numerical results from the continuum limit is due to the numerical viscosity (or dissipation) terms introduced by the discretization scheme in order to increase its precision. 
There is no accompanying noise for this dissipative channel.
These additional contributions are absent in the continuum equations and correspond to the off-diagonal terms in Eqs.~(\ref{predictor_step}-\ref{corrector_step}). In the predictor step, they have the form
$
-c_s(2\delta\varepsilon_{i}^n-\delta\varepsilon_{i+1}^n-\delta\varepsilon_{i-1}^n)
$
and
$
-c_s(2\delta g_{i}^n-\delta g_{i+1}^n-\delta g_{i-1}^n)
$
with analogous terms appearing in the corrector step.

\begin{figure}
    \centering
    \includegraphics[width=\linewidth]{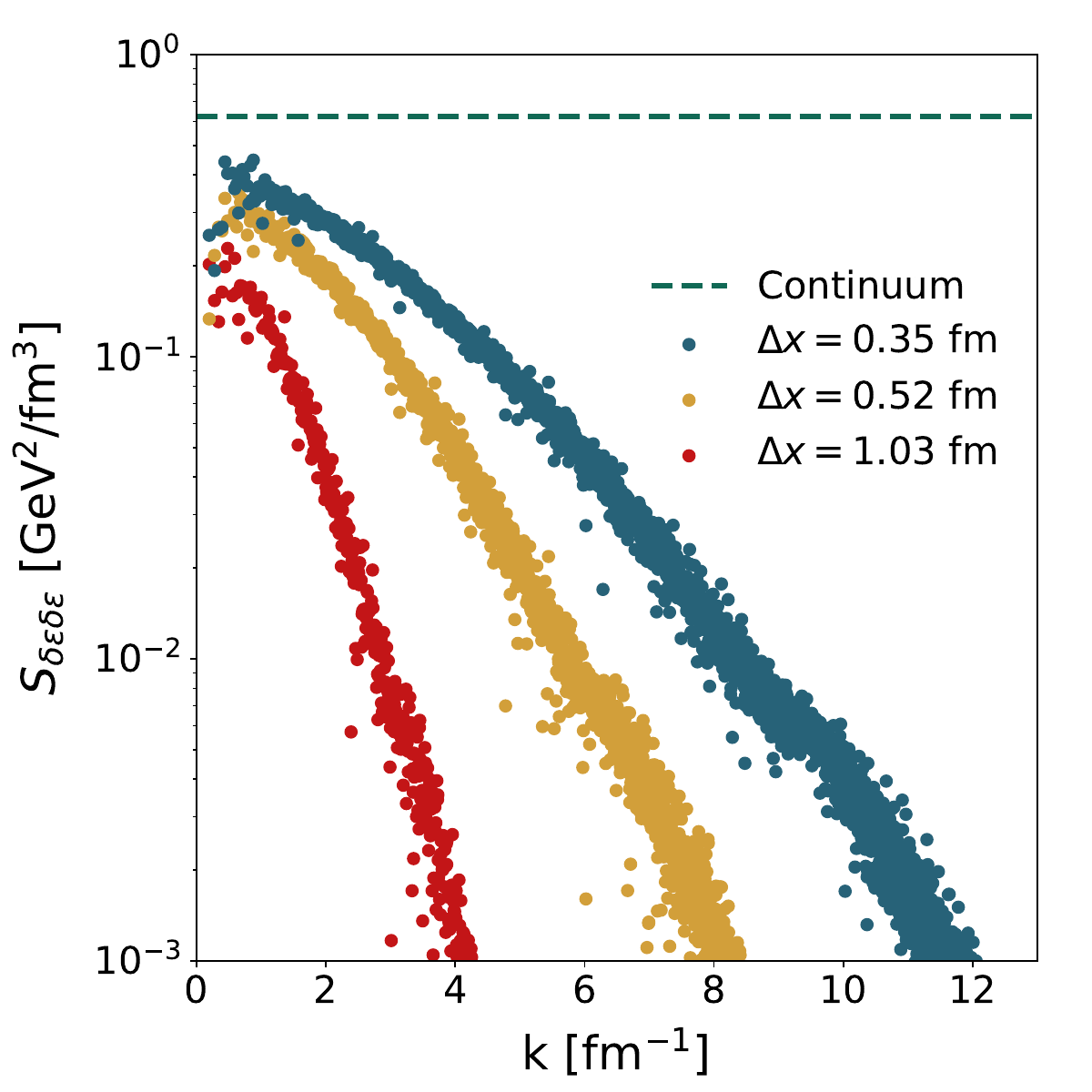}
    \caption{Three-dimensional Israel-Stewart  static structure factor for energy density. The analytical continuum result according to Eq.~(\ref{e:anal_SF}) with green dashed line.  Data points show results from simulations with linear energy density profile within cells, $\eta/s = 0.12$, and time step $\Delta t = 0.1$~fm/$c$, for three different cell sizes.}
    \label{fig:3D_linear}
\end{figure}

The agreement of numerical and analytical results in Fig.~\ref{fig:1D_flat} is important as it indicates that there are no additional numerical artifacts that would have escaped our understanding. 
With this knowledge we will now move on to the three-dimensional simulations where we do not have the analytical results to compare to. 
Also, we now allow the fields to change linearly within the cells (which is also known as piecewise linear distributions), since this more precise scheme is the default choice in the algorithm.
The static structure factors as  functions of $k = |\vec k|$ for three different cell sizes are plotted in Fig.~\ref{fig:3D_linear}.
Qualitatively, the same feature is observed as in the 1D simulations: for larger cell sizes the structure factors differ from the continuum result more. 
As before, the difference between the continuum result and simulations is always quite large.

\begin{figure}
    \centering
    \includegraphics[width=\linewidth]{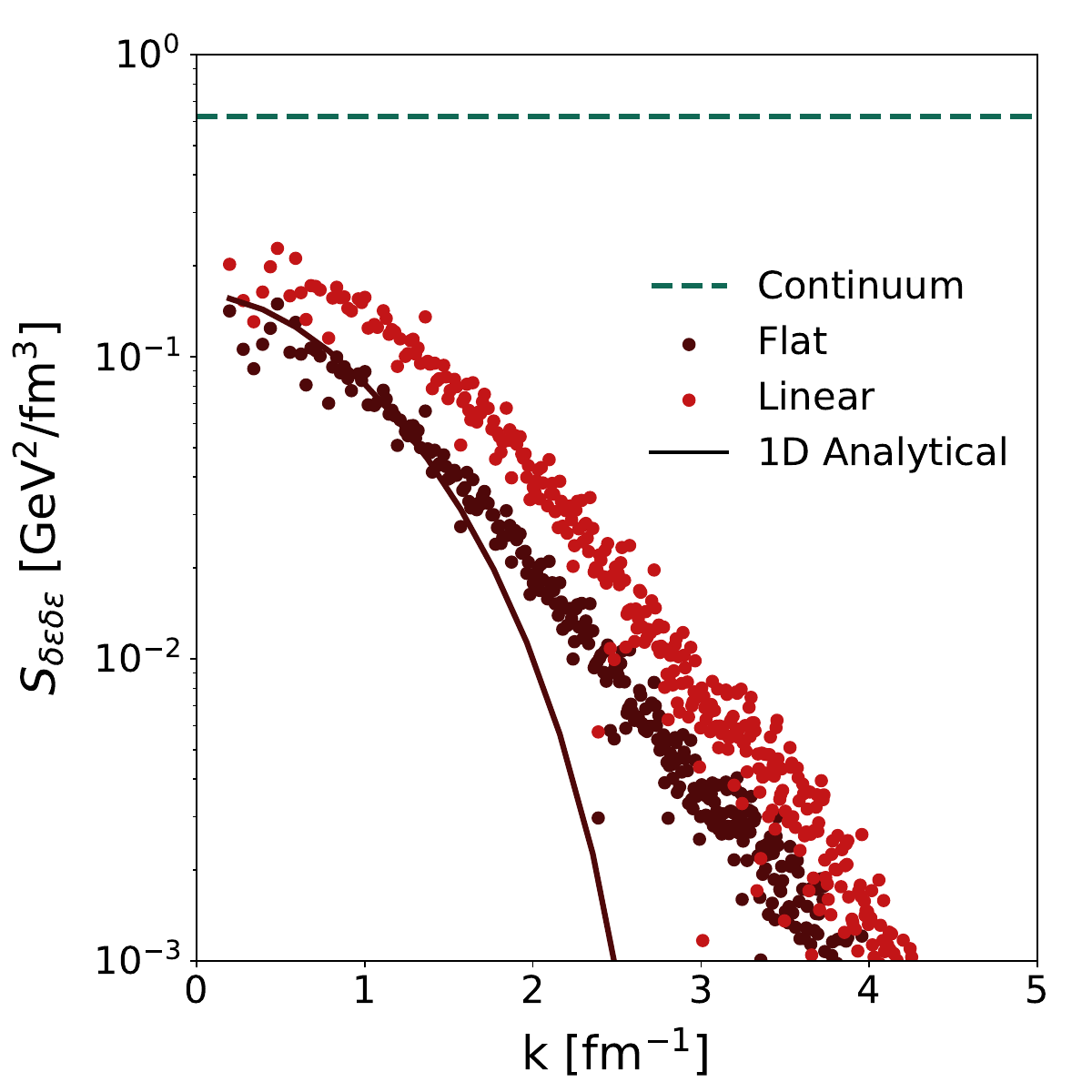}
    \caption{Comparison of static structure factors for energy density from 3D simulations with piecewise constant cell profiles (burgundy) and piecewise linear cell profiles (red). The full line represents 1D semi-analytic calculation for piecewise constant cell profiles
    and the dashed line the corresponding continuum limit result. 
    Simulations with $\Delta t = 0.1$~fm/$c$, $\Delta x = 1.03$~fm, and $\eta/s = 0.12$.}
    \label{fig:3D_flat_vs_linear}
\end{figure}

We compare the schemes for the numerical solution of the IS equations in Fig.~\ref{fig:3D_flat_vs_linear}.
Here, 3D numerical simulation with piecewise constant distributions of the fields within cells---as explained above---is compared with the piecewise linear distributions. 
The latter is the default choice in vHLLE code.
Piecewise linear distributions which improve the accuracy of regular numerical solutions, yield a result that is closer to the continuum limit, even though the deviation is still quite large. 
It is also shown that the semi-analytic 1D solution describes the result from 3D simulations with piecewise constant distributions within cells rather well at small $k$, though it eventually  differs by several orders of magnitude for large $k$. 

The structure factor measures spatial correlations of the fluctuations, hence it is intuitively conceivable that its numerical solution will depend on $\Delta x$.
Even though it is not related to the temporal correlation, the formulae (\ref{e:ab:Mk}) and (\ref{e:ab:Qk}) include terms with $\Delta t$, so that one might expect a dependence on the time step, as well.
We have checked, however, that such a dependence does not appear in the numerical results.

\begin{figure}
    \centering
    \includegraphics[width=\linewidth]{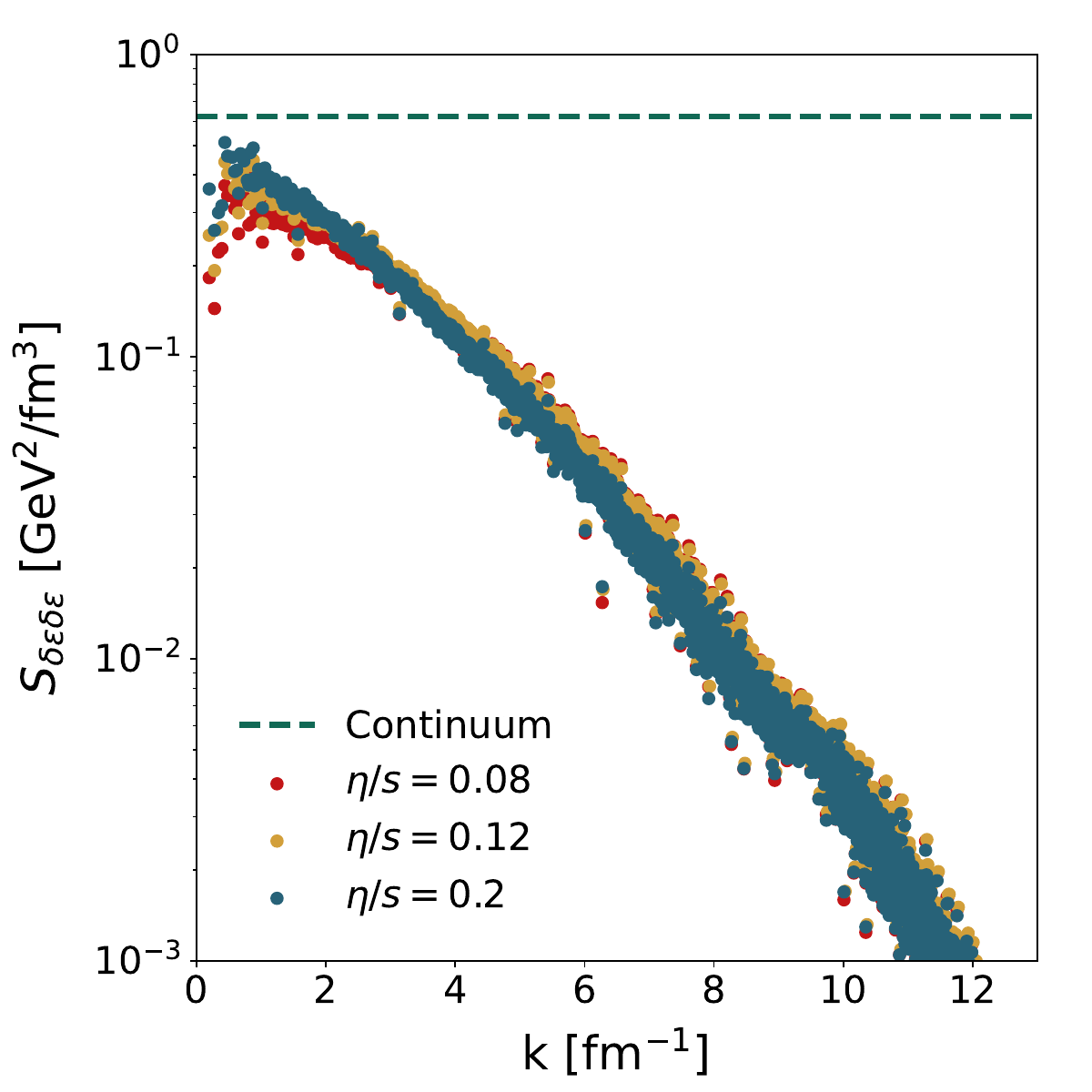}
    \caption{Three-dimensional Israel-Stewart  static structure factor for energy density. The analytical continuum result according to Eq.~(\ref{e:anal_SF}) with green dashed line.  Data points show results from simulations with piecewise linear energy density profile within cells for viscosity $\eta/s = 0.08$, $\eta/s = 0.12$ and $\eta/s = 0.2$, time step $\Delta t = 0.1$~fm/$c$, for cell size $\Delta x=0.35$ fm.}
    \label{fig:3D_linear_eta}
\end{figure}

Let us note that the continuum static structure factors for all correlators  are not only independent of $k$, but also of the shear viscosity, as expected for equilibrium quantities, see Eq.~(\ref{e:anal_SF}). 
This may be not obvious for the $\delta\pi\delta\pi$ correlator, but recall that $\tau_\pi$ scales with $\eta$ (cf.~Eq.(\ref{e:tau-visc})), hence the $\eta$-dependence scales out.
Viscosity appears in the matrix equations in the derivation of the structure factor for the discretised space but in the end we invert those equations  numerically. 
Therefore, we study the influence of the viscosity on the  $\delta\varepsilon \delta\varepsilon$ structure factor in simulations and show their results in Fig.~\ref{fig:3D_linear_eta}.
For all examined values of $\eta/s$ the structure factor decays with the same width. 
The maximum of the structure factor at small $k$ is higher and closer to the continuum result in cases of larger viscosity.


\section{Viscosity from fluctuations}
\label{s:visc}

In fluid dynamics, shear viscosity quantifies the shear stress that corresponds to momentum transfer between the neighbouring layers of fluid which move with different velocities. As such, the shear stress tensor is proportional to the gradients of collective flow velocity $\vec v$.
As for the fluctuations, their amplitudes have been set by the fluctuation-dissipation theorem. 
It is the dissipation, parametrised by viscosity, that determines the correlation between the shear stresses at different times. 
This is reflected by the Green-Kubo relation
\begin{equation}
    \label{e:GK_xy}
    \eta_f=\frac{1}{T}
    \int\textrm{d}^3\vec r\int_0^\infty\textrm{d}t\langle\delta\pi^{\prime xy}(0,0)\delta\pi^{\prime xy}(\vec r,t))\rangle\, .
\end{equation}
The subscript $f$ in $\eta_f$ indicates that this is the viscosity determined from the correlations of fluctuations, as for the moment we want to distinguish it from the input $\eta$ that enters the IS equations. 
Since fluctuations  neither depend on time nor on position, part of the ensemble averaging can be technically turned into the averaging over the volume. We thus introduce the averaged transverse shear stress fluctuation
\begin{equation}
    \delta \Pi^{xy}(t) = \frac{1}{V} 
    \int \textrm{d}^3\vec r \, \delta\pi^{\prime xy}(\vec r,t)\, ,
\end{equation}
and rewrite Eq.~(\ref{e:GK_xy}) as
\begin{equation}
    \eta_f = \frac{V}{T} \int_0^\infty \mathrm{d}t
    \langle \delta \Pi^{xy}(t_0)\delta \Pi^{xy}(t_0+t)\rangle \, .
\label{e:GK_xy_eval}    
\end{equation}

Before embarking on numerical evaluation of the correlator under the integral, let us analyze its behaviour. 
By integrating Eq.~(\ref{e:v0delpi}) over the volume we obtain
\begin{equation}
    \partial_t \delta \Pi^{xy}  = - \frac{\delta \Pi^{xy}}{\tau_{\pi,0}} + 
    \frac{\int_V \delta \pi_{NS}^{xy} \, dV}{\tau_{\pi,0}}
    + \frac{\Psi^{xy}}{\tau_{\pi,0}}\, ,
\end{equation}
where $\Psi^{xy}$ is volume-integrated white noise. 
The integral in the second term vanishes since $\delta \pi_{NS} \propto \partial \delta u$ and the volume integral is the $k=0$ mode.
The equation reduces to Ornstein-Uhlenbeck process and can be written as
\[
\frac{d\phantom{t}}{dt} \left [ e^{t/\tau_{\pi,0}} \delta \Pi^{xy}(t)\right ]\
= \frac{1}{\tau_{\pi,0}} e^{t/\tau_{\pi,0}} \Psi^{xy}(t)\, .
\]
Integrating this equation from $-\infty$ to $t$ gives
\[
\delta \Pi^{xy} = \frac{1}{\tau_{\pi,0}} 
\int_{-\infty}^t ds\, e^{-(t-s)/\tau_{\pi,0}} \Psi^{xy}(s)\, .
\]
The correlation of the volume-averaged noise is given by the fluctuation-dissipation relation
\begin{equation}
\left \langle   \Psi^{xy}(t) \Psi^{xy}(t')\right \rangle = 
\frac{2\eta_0 T_0}{V} \delta (t-t')\, .
\end{equation}
Using this we arrive at the variance
\begin{multline}
\left \langle \delta\Pi^{xy}(t) \delta\Pi^{xy}(t')\right \rangle =\\
\frac{1}{\tau_{\pi,0}^2} \int_{-\infty}^t ds \int_{-\infty}^{t'} dr 
e^{(t-s)/\tau_{\pi,0}} e^{(t'-r)/\tau_{\pi,0}} 
\\ \times
\left \langle   \Psi^{xy}(s) \Psi^{xy}(r)\right \rangle
= \\
\frac{\eta_0 T_0}{\tau_{\pi,0}V} \exp\left ( - \frac{(t-t')}{\tau_{\pi,0}}\right )\, 
\label{e:corPiPi}
\end{multline}
This shows that ${\eta_0 T_0}/{\tau_{\pi,0}V}$ is the stationary variance of the Ornstein-Uhlenbeck process and the correlation decays exponentially at the rate given by $\tau_{\pi,0}$.

Note that plugging  Eq.~(\ref{e:corPiPi}) directly into Eq.~(\ref{e:GK_xy_eval}) and performing the integral there yields $\eta_f = \eta_0$, since $T=T_0$.

We have numerically evaluated the correlator under the integral in Eq~(\ref{e:GK_xy_eval}). 
For the ensemble averaging, the entries are separated by a minimum delay to avoid correlations. 
The correlator actually resulting from the simulations is rather fluctuating as function of time due to limited statistics available. 
This made the direct integration according to Eq.~(\ref{e:GK_xy_eval}) numerically ambiguous. 
Since according to (\ref{e:corPiPi}), the correlator is expected to  decrease exponentially, we fit it with the parametrisation
\begin{equation}
   \langle \delta \Pi^{xy}(t_0)\delta \Pi^{xy}(t_0+t)\rangle
   = A e^{-t/\tau_{\mathrm{rel}}}\, , 
\end{equation}
where the fit parameters are 
\begin{equation}
    A = \langle \delta \Pi^{xy}(0)\delta \Pi^{xy}(0)\rangle
\end{equation}
and $\tau_{\mathrm{rel}}$. 
We verified a posteriori that $\tau_{\mathrm{rel}}$ agrees numerically with $\tau_{\pi,0}$, as expected from Eq.~(\ref{e:corPiPi}). 

With the fitted exponential parametrisation the integral in Eq.~(\ref{e:GK_xy_eval}) is performed easily and yields
\begin{equation}
    \eta_f = \frac{V \tau_{\mathrm{rel}}}{T} 
    \langle \delta \Pi^{xy}(0)\delta \Pi^{xy}(0)\rangle\, .
\end{equation}

The uncertainty interval on $\eta_f$ is obtained from performing the fit on 10 independent  consecutive time blocks of the simulation and calculating the mean and variance from the ensemble of the obtained $\eta_f$'s from all blocks. 

\begin{figure}
    \centering
    \includegraphics[width=\linewidth]{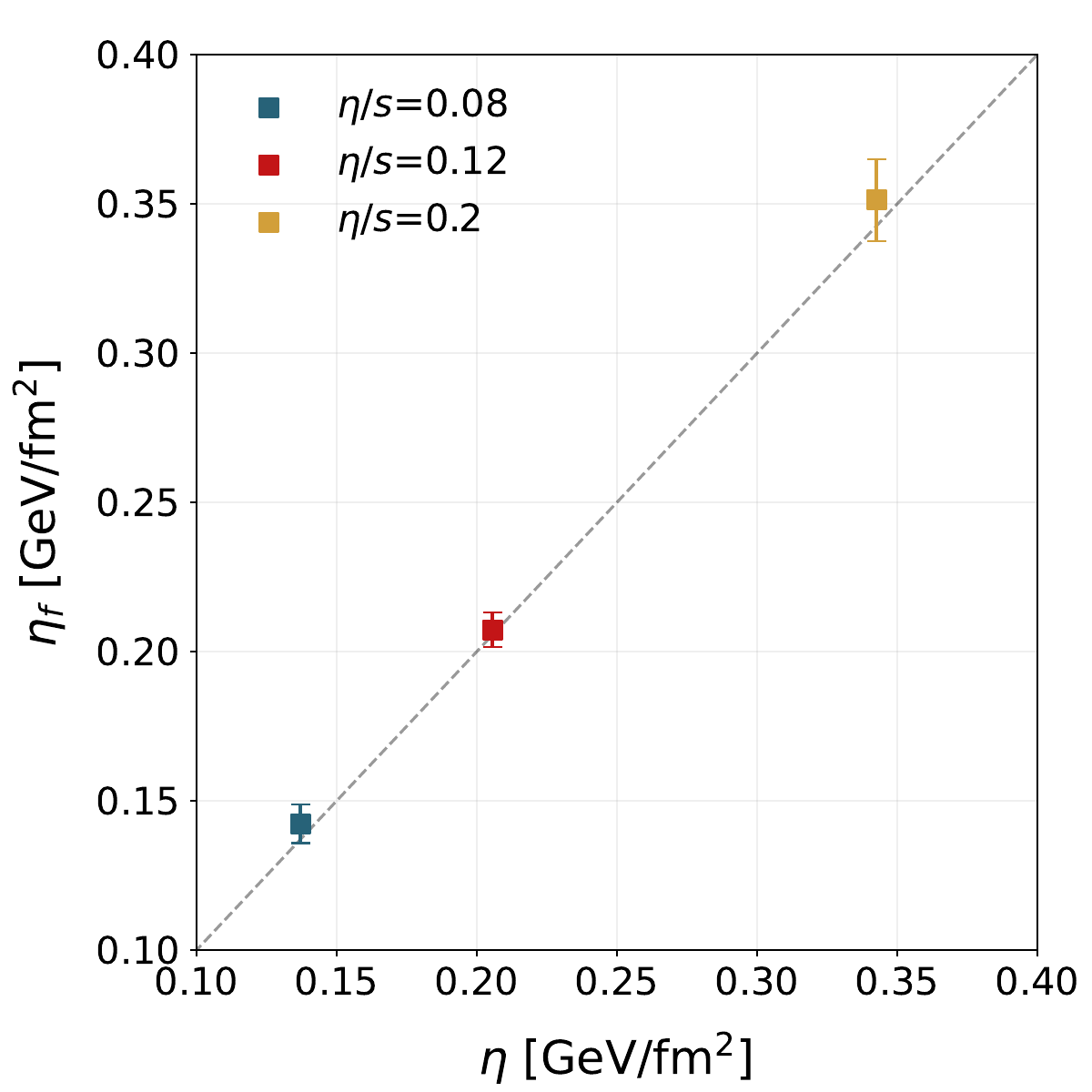}
    \caption{Dependence of the fitted values $\eta_f$ according 
    to Eq.~(\ref{e:GK_xy_eval}) on the shear viscosity $\eta$ entering the IS equations.}
    \label{f:eta-eta}
\end{figure}

Our main result is shown in Fig.~\ref{f:eta-eta}.
We have performed the simulations for three different values of $\eta$ (expressed through $\eta/s$) as we want to see that $\eta_f$ obtained from Eq.~(\ref{e:GK_xy}) indeed gives the same value. 
The extracted values agree with the input within the uncertainties.
This validates the treatment of linearised fluctuations. 
To our knowledge, this is the first extraction of viscosity from fluctuations that has been performed  for relativistic fluid dynamics. 

Since there is an explicit volume factor  in Eq.~(\ref{e:GK_xy_eval}) and the amplitude of fluctuations itself depends on the cell size, we have checked that the resulting $\eta_f$ depends on neither of them. 
\begin{figure}
    \centering
    \includegraphics[width=\linewidth]{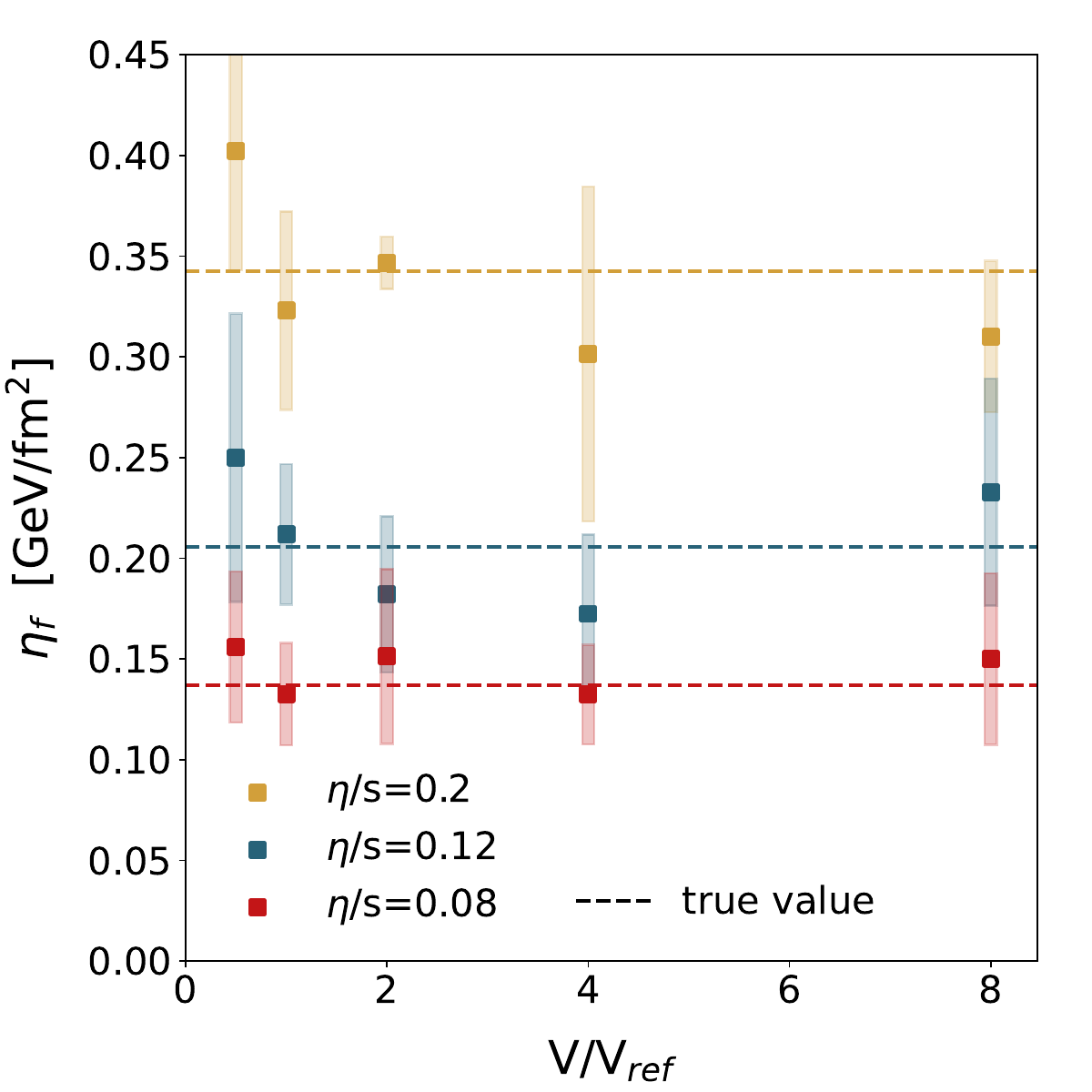}
    \includegraphics[width=\linewidth]{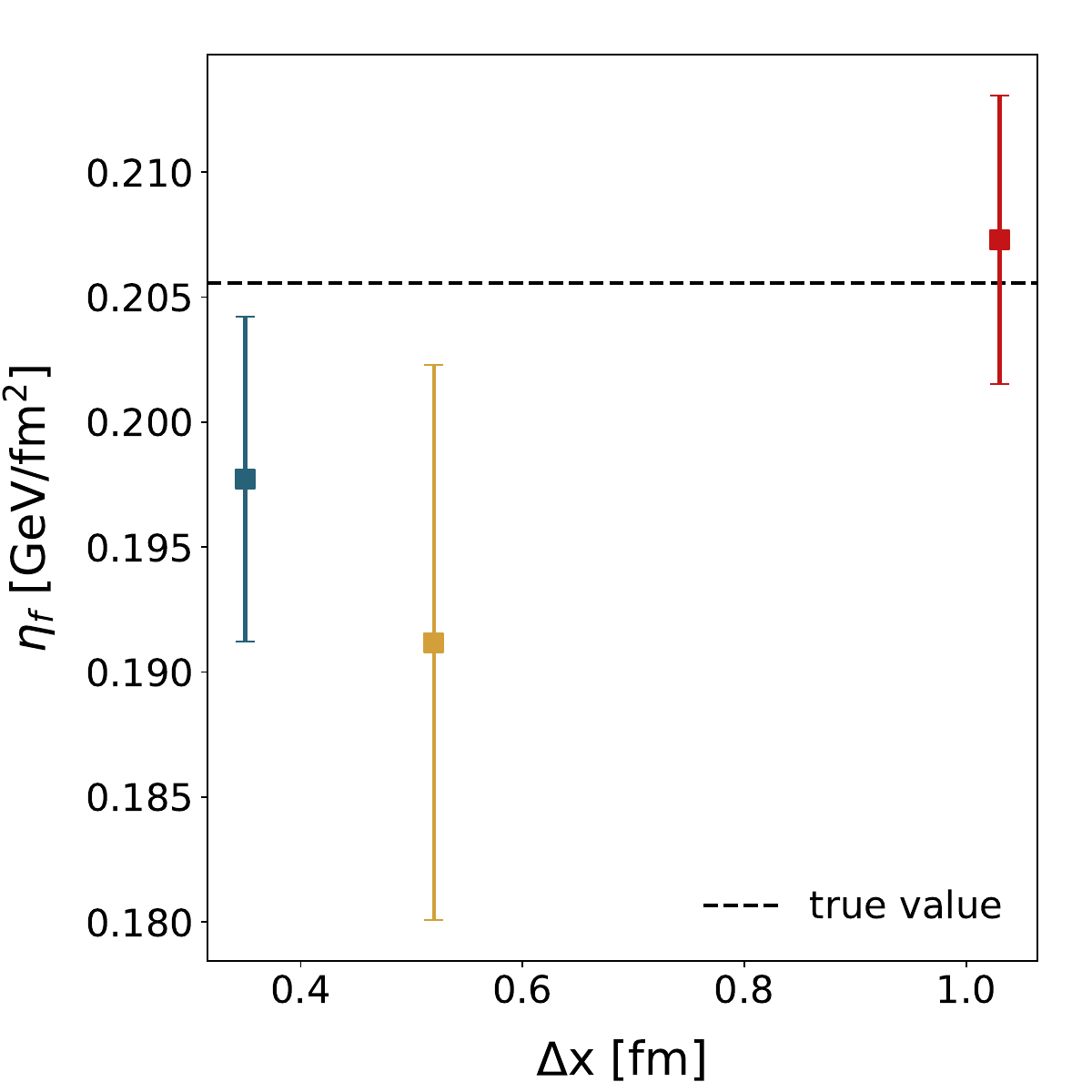}
    \caption{Top: A plot of $\eta_f$ as a function of $V/V_{\mathrm{ref}}$, where the reference volume $V_{\mathrm{ref}}$ is 32$^3$~fm$^3$, for which usually our calculations are done. 
    Different symbols represent different $\eta/s$. 
    Bottom: $\eta_f$ for $\eta/s = 0.12$ plotted as function of cell size.}
    \label{fig:eta_V}
\end{figure}
This is demonstrated in Fig.~\ref{fig:eta_V}. 
Within the errorbars, the upper panel shows that the extracted $\eta_f$ does not depend on the total volume in a range where the volume changes up to 16 times.
In the lower panel, for fixed $\eta/s$ we measure $\eta_f$ for three different values of the cell size $\Delta x$ and observe no dependence on it within the uncertainty intervals. 

\subsection{Impact of the viscous limiter in vHLLE}

\begin{figure}
    \centering
    \includegraphics[width=\linewidth]{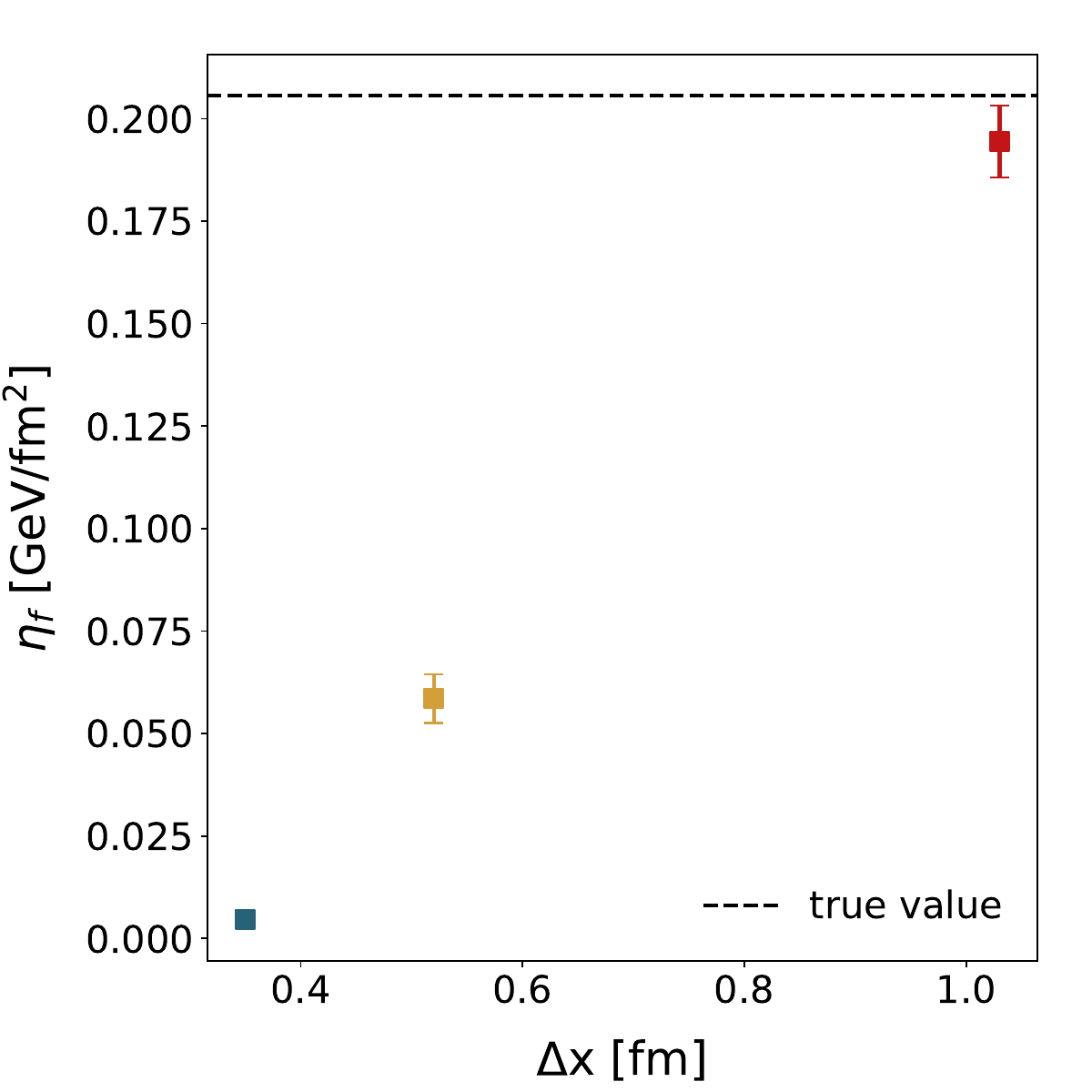}
    \caption{The extracted $\eta_f$ as function of cell size for the cases where limiter is invoked. 
    Dashed line shows the value of $\eta$ on input.}
    \label{f:limiter}
\end{figure}

Note that the above simulations for the extraction of viscosity were performed with the fluctuations scaled by $\alpha = 0.01$.
Since this cannot be done for all observables, fluctuations may in some cases be so large that their contributions need to be cut down by a limiter. 
As can be seen from Eq.~(\ref{e:xipiwidth}), this is more likely for smaller cells, as the amplitude of the noise is inversely proportional to the square root of the cell volume. 
We look at the impact of the limiter in Fig.~\ref{f:limiter}.
The extracted $\eta_f$ is plotted for $\eta/s = 0.12$ as a function of the cell size $\Delta x$. 
The limiter was invoked in 0.03\%, 12.3\%, and 38.5\% of all cell updates, for $\Delta x=1.03$~fm, 0.52~fm, and 0.35~fm, respectively.
The effect of the limiter is dramatic. 
For the smallest cell size the time correlator is almost completely destroyed by the limiter and the obtained $\eta_f$ is almost vanishing.


\section{Conclusions}
\label{s:conc}

An accurate treatment of thermal fluctuations in relativistic fluid-dynamical simulations is particularly important in the search for the critical point of the QCD phase diagram, where fluctuations are expected to be enhanced.
Thermal fluctuations may also be relevant at higher collision energies, where they provide an additional source of event-by-event fluctuations on top of those originating from the initial state and may contribute to final-state flow observables.

In this work, we formulated and studied one possible implementation of thermal fluctuations in relativistic viscous hydrodynamics, in which they are treated as linear perturbations on top of a smooth averaged background.
In order to benchmark the stochastic implementation against analytical results, we considered a homogeneous static background in a box with periodic boundary conditions.
This setup allows us to study the properties of the fluctuating sector independently of the evolution of the background.

We investigated the spatial correlations of the fluctuations using static structure factors.
For the continuum linearized theory considered here, the equilibrium structure factors of the energy-density and velocity fluctuations are independent of the wave number and of the shear viscosity.
The numerical discretisation changes this result.
At finite grid spacing, the structure factors acquire a non-trivial wave-number dependence and deviate increasingly from the continuum prediction towards shorter wavelengths.
We also find a residual, albeit  smaller, dependence on the shear viscosity, which is absent from the continuum equilibrium result and therefore provides an additional measure of finite-resolution effects.
The structure factor thus offers a sensitive test of whether a numerical stochastic hydrodynamic scheme reproduces the equilibrium fluctuations required by the fluctuation-dissipation relation.

To probe the dynamics of the fluctuations, we studied the temporal correlation of the shear stress and its relation to the shear viscosity through the Green--Kubo formula \cite{Kubo:1966fyg,Hosoya:1983id}.
When the stochastic evolution is not modified by the viscous regulator, the viscosity extracted from the shear-stress correlation function agrees with the viscosity used as input to the hydrodynamic evolution.
This provides a direct numerical consistency check of the fluctuation-dissipation relation in the relativistic fluid-dynamical framework considered here.
The situation changes when the viscous regulator becomes active. As the grid is refined, the stochastic contributions to the shear stress become increasingly large, and therefore the viscous stress is regulated to preserve stability of the hydrodynamic evolution. This regulation is not accompanied by the appropriate changes in the noise correlation and therefore breaks the fluctuation-dissipation relation.
We find that this can lead to a strong violation of the fluctuation-dissipation relation: in particular, the viscosity obtained from the Green--Kubo relation no longer reproduces the input value.

These observations distinguish two numerical issues which are often entangled in simulations with hydrodynamic fluctuations.
Even for a coarse resolution, where the evolution remains in the linear regime and the viscous regulator is inactive, the finite spatial and temporal discretization modifies the equilibrium fluctuation spectrum.
When the stochastic stresses become sufficiently large for the regulator to set in, an additional and considerably stronger modification of the fluctuation-dissipation balance occurs.
Both effects become increasingly relevant when shorter-wavelength fluctuations are resolved.

Our results therefore show that reproducing the noise correlator locally is not by itself sufficient for a consistent numerical implementation of fluctuating hydrodynamics.
The equilibrium correlations generated by the complete numerical evolution have to be tested as well.
Static structure factors and Green--Kubo relations provide complementary benchmarks for this purpose.
For applications to realistic heavy-ion collisions, where the background is rapidly evolving and stochastic fluctuations coexist with stability requirements of viscous hydrodynamics, controlling the treatment of short-wavelength fluctuations and establishing the range of wavelengths over which the numerical evolution reproduces the fluctuation-dissipation relation will be important steps towards a quantitatively reliable implementation.

\begin{acknowledgements}
We thank Marcus Bluhm for valuable discussions on calculations of the structure factor. 
JS, IK, and BT acknowledge the support from the Czech Science Foundation, under No 25-16877S. 
The collaboration on this project was supported by a Barrande fellowship awarded to JS.
\end{acknowledgements}


\appendix

\section{Symbols and definitions in the linearised hydrodynamic equations}
\label{a:symbols}

The transport coefficients for the evolution of shear and bulk stress are
\begin{equation}
\begin{split}
    \delta_{\pi\pi}=\frac{4}{3}\tau_\pi,\hspace{0.5cm}\tau_{\pi\pi}&=\frac{10}{7}\tau_\pi,\hspace{0.5cm}\phi_7=\frac{9}{70p}\\
    \lambda_{\pi\Pi} = \frac{6}{5}\tau_\pi,\hspace{0.5cm}\delta_{\Pi\Pi} &= \frac{2}{3}\tau_\Pi,\hspace{0.5cm}\lambda_{\Pi\pi} = \frac{8}{5}\bigg(\frac{1}{3} -c_s^2(e)\bigg)\tau_\Pi
\end{split}.
\label{e:coefs}
\end{equation}
The Navier-Stokes values of the shear stress and bulk pressure for the background are calculated with the projector to the background velocity
\begin{eqnarray}
\label{e:shear_NS}
\pi^{\mu\nu}_{0,NS} & = & 2\eta_0 \sigma_0^{\mu\nu} 
\\ \nonumber & = & 
\eta_0 \left ( \Delta_0^{\mu\lambda} \partial_\lambda u^\nu_0 + 
\Delta_0^{\nu\lambda} \partial_\lambda u^\mu_0 \right ) - \frac{2}{3} \eta_0 \Delta_0^{\mu\nu}\partial_\lambda u_0^\lambda \\
\Pi_{0,NS} & = & - \zeta_0 \partial_\mu u_0^\mu 
\end{eqnarray}

The shear is decomposed in Eqs. (\ref{e:IS_fluc}) into 
\begin{eqnarray}
\sigma_0^{\mu\nu} & = & \frac{1}{2} \frac{\pi_{NS}^{\mu\nu}}{\eta_0}\\
\delta\sigma^{\mu\nu} & = & 
\frac{1}{2} \frac{\pi_{NS}^{\mu\nu} - 2\delta\eta\, \sigma_0^{\mu\nu}}{\eta_0}
\end{eqnarray}

The fluctuating part of the NS shear stress is identified as
\begin{multline}
\delta \pi_{NS}^{\mu\nu} = \eta_0 \Bigg [
\left [ 
\Delta_0^{\mu\lambda} \partial_\lambda \delta u^\nu
- ( \delta u^\mu u_0^\lambda + u_0^\mu \delta u^\lambda )\partial_\lambda u_0^\nu
\right ]
\\
+ \left [ 
\Delta_0^{\nu\lambda} \partial_\lambda \delta u^\mu
- ( \delta u^\nu u_0^\lambda + u_0^\nu \delta u^\lambda )\partial_\lambda u_0^\mu
\right ]
\\
- \frac{2}{3}
\left [ 
\Delta_0^{\mu\nu} \partial_\lambda \delta u^\lambda
- ( \delta u^\mu u_0^\nu + u_0^\mu \delta u^\nu )\partial_\lambda u_0^\lambda
\right ]
\Bigg ]
\\
+ \delta \eta
\left [
\Delta_0^{\mu\lambda}\partial_\lambda u_0^\nu + \Delta_0^{\nu\lambda}\partial_\lambda u_0^\mu
- \frac{2}{3} \Delta_0^{\mu\nu} \partial_\lambda u_0^\lambda
\right ]\, ,
\end{multline}
\begin{equation}
\delta \Pi_{NS} = -\zeta_0 \partial_\mu \delta u_0^\mu 
- \delta\zeta\, \partial_\mu u_0^\mu\, .
\end{equation}


\section{Analytical calculations for the static structure factor in 1D}
\label{a:anal-SF}

\subsection{Analytical calculation in continuum}
\label{a:anal-sf-cont}

In general, the equations for relativistic viscous fluid dynamics take the  form
\begin{equation}
    \partial_tU=LU+KW
\end{equation}
where $U$ is a vector of evolved fields, $L$ is a linear evolution operator, $K$ is noise coupling operator and $W$ is stochastic noise. In Israel-Stewart framework, the equations take particular form

\begin{subequations}
\label{e:a:block}
\begin{eqnarray}
\label{IS_eq_set}
    \partial_t\delta\varepsilon&=&-(\varepsilon_0+p_0)\partial_x\delta u\\
    \partial_t\delta u&=&-\frac{c_s^2}{\varepsilon_0+p_0}\partial_x\delta\varepsilon-\frac{1}{\varepsilon_0+p_0}\partial_x \delta\pi\\
    \partial_t\delta\pi&=&-\frac{\delta\pi}{\tau_{\pi,0}}+\frac{4}{3}\frac{\eta_0}{\tau_{\pi,0}}\partial_x\delta u + \frac{\xi}{\tau_{\pi,0}}
\end{eqnarray}
\end{subequations}
Here, the vector $U$ consists of the field of energy density $\delta\varepsilon$, velocity $\delta u$ and shear-stress $\delta\pi$
\begin{equation}
U = \left (
\begin{array}{c}
\delta\varepsilon \\ \delta u \\ \delta \pi
\end{array}
\right )\, .
\end{equation} 
The noise is denoted by $\xi$. For easier manipulation, let us rewrite $\varepsilon_0+p_0$ as enthalpy $w_0$.\\
\indent To calculate the structure factor $S$, we must solve Lyapunov equation \cite{Donev:2010}
\begin{equation}
\label{lyapunov}
    L_kS+SL_k^\dagger=-K_kC_WK_k^\dagger
\end{equation}
where the index $k$ denotes Fourier transform and the dagger stands for Hermitian conjugate.

From (\ref{e:a:block}) we can identify the operator $L$ as
\begin{equation}
    L=
    \begin{pmatrix}
        0 & -w_0\partial_x & 0\\
        -\frac{c_s^2}{w_0}\partial_x & 0 & -\frac{1}{w_0}\partial_x\\
        0 & -\frac{4}{3}\frac{\eta_0}{\tau_{\pi,0}}\partial_x & -\frac{1}{\tau_{\pi,0}}
    \end{pmatrix}
\end{equation}
and its Fourier transform
\begin{equation}
    L_k=
    \begin{pmatrix}
        0 & -ikw_0 & 0\\
        -ik\frac{c_s^2}{w_0} & 0 & -ik\frac{1}{w_0}\\
        0 & -ik\frac{4}{3}\frac{\eta_0}{\tau_{\pi,0}} & -\frac{1}{\tau_{\pi,0}}
    \end{pmatrix}
\end{equation}
\begin{equation}
    L_k^\dagger=
    \begin{pmatrix}
        0 & ik\frac{c_s^2}{w_0}& 0\\
        ikw_0 & 0 & ik\frac{4}{3}\frac{\eta_0}{\tau_{\pi,0}}\\
        0 & ik\frac{1}{w_0} & -\frac{1}{\tau_{\pi,0}}
    \end{pmatrix} \, .
\end{equation}
Similarly, we identify $K$
\begin{equation}
    K=
    \begin{pmatrix}
        0 & 0 & 0\\
        0 & 0 & 0\\
        0 & 0 & \frac{1}{\tau_{\pi,0}}
    \end{pmatrix}=K_k=K_k^\dagger\, .
\end{equation}
Finally, the noise covariance matrix $C_W$ is set in accordance with Eq. (\ref{e:xicor})
\begin{equation}
    C_W= \begin{pmatrix} 0 &0 & 0\\ 
    0 & 0 & 0 \\ 0 & 0 &\frac{8}{3}\eta_0 T_0 \end{pmatrix}\, .
\end{equation}
Putting all into the Eq. (\ref{lyapunov}) and under the assumption that all off-diagonal elements of $S$ are zero, 
the resulting static structure factor matrix in 1D is
\begin{equation}
\label{e:anal_SF_IS}
    S_k=
    \begin{pmatrix}
        \frac{(\varepsilon_0+p_0)T_0}{c_s^2} & 0&0\\
        0 & \frac{T_0}{\varepsilon_0+p_0} & 0\\
        0&0& \frac{4}{3}\frac{\eta_0 T_0}{\tau_{\pi,0}}
    \end{pmatrix}\, .
\end{equation}

\subsection{Semi-analytical calculation for vHLLE discretized scheme in 1D}
\label{a:sol1d}

Generally, explicit linear finite-volume schemes for solving the stochastic partial differential equations (SPDE) are specified by a linear recursion of the form
\begin{equation}
    X^{n+1}=MX^n+NW^n
\end{equation}
where $W$ is a vector of standard normal variables with mean zero and variance one, and the superscript denotes time step.  
In order to calculate the structure factor matrix $S$, we need to solve discrete time Lyapunov equation
\begin{equation}
    M_kSM_k^\dagger-S= -\Delta xN_kC_WN^\dagger_k\, ,
\end{equation}
where $M_k$ and $N_k$ are Fourier transformed $M$ and $N$, respectively,  $C_W$ is the correlation matrix of the white noise, equal to identity, and $\Delta x$ is the size of the elementary cell.
We denote the r.h.s.\ of this equation as 
\begin{equation}
    Q_k = -\Delta xN_kC_WN^\dagger_k\, .
\end{equation}
We consider a stochastic hydrodynamic system in one spatial dimension, where the conserved variables are the energy density perturbation $\delta \varepsilon$ and momentum density perturbation $\delta g = w_0 \delta u$, with $w_0 = \varepsilon_0 + p_0$ the background enthalpy. In the Israel-Stewart (IS) theory, the shear stress $\delta \pi$ is promoted to an independent dynamical variable. The state vector is therefore
\begin{equation}
    X_i^n =
    \begin{pmatrix}
        \delta \varepsilon_i^n \\
        \delta g_i^n \\
        \delta \pi_i^n
    \end{pmatrix}\, ,
\end{equation}
where the subscript $i$ denotes discretized spatial position. 

The time evolution is implemented using a predictor--corrector scheme. 
Recall that we work in 1D.

The ideal hydrodynamic variables are updated in two steps. 
First step
\begin{widetext}
\begin{equation}
\label{predictor_step}
    \begin{split}
    \delta\varepsilon_{i}^{n+\tfrac{1}{2}}&=\delta\varepsilon_{i}^n+\frac{1}{2}\frac{\Delta t}{2\Delta x}\bigg[ \delta g_{i-1}^{n}-\delta g_{i+1}^{n}-c_s\Big(2\delta\varepsilon_{i}^n-\delta\varepsilon_{i+1}^n-\delta\varepsilon_{i-1}^n\Big) \bigg]\\
    \delta g_{i}^{n+\tfrac{1}{2}}&=\delta g_{i}^{n}+\frac{1}{2}\frac{\Delta t}{2\Delta x}\bigg[ c_s^2\Big(\delta \varepsilon_{i-1}^{n}-\delta \varepsilon_{i+1}^{n}\Big)-c_s\Big(2\delta g_{i}^{n}-\delta g_{i+1}^{n}-\delta g_{i-1}^{n}\Big)\bigg]
    \end{split}
\end{equation}
and the second step which uses the result of the first step
\begin{equation}
\label{corrector_step}
    \begin{split}
    \delta\varepsilon_{i}^{n+1*}&=\delta\varepsilon_{i}^n+\frac{1}{2}\frac{\Delta t}{\Delta x}\bigg[ \delta g_{i-1}^{n+\tfrac{1}{2}}-\delta g_{i+1}^{n+\tfrac{1}{2}}-c_s\Big(2\delta\varepsilon_{i}^{n+\tfrac{1}{2}}-\delta\varepsilon_{i+1}^{n+\tfrac{1}{2}}-\delta\varepsilon_{i-1}^{n+\tfrac{1}{2}}\Big) \bigg]\\
    \delta g_{i}^{n+1*}&=\delta g_{i}^{n}+\frac{1}{2}\frac{\Delta t}{\Delta x}\bigg[ c_s^2\Big(\delta \varepsilon_{i-1}^{n+\tfrac{1}{2}}-\delta \varepsilon_{i+1}^{n+\tfrac{1}{2}}\Big)-c_s\Big(2\delta g_{i}^{n+\tfrac{1}{2}}-\delta g_{i+1}^{n+\tfrac{1}{2}}-\delta g_{i-1}^{n+\tfrac{1}{2}}\Big) \bigg]\, .
    \end{split}
\end{equation}
The star in the superscript indicates that there will be one more update in which the shear stress will be accounted for. 
The shear stress evolves via a relaxation equation with stochastic forcing, also in two steps
\begin{equation}
    \begin{split}
    \delta\pi_i^{n+\frac{1}{2}}=\delta\pi_i^{n}-\frac{\Delta t}{2\gamma\tau_{\pi_0}}\Big( \delta\pi_i^n-\delta\pi_{NS,i}^{n}-\xi_i^n \Big)\\
    \delta\pi_i^{n+1}=\delta\pi_i^{n}-\frac{\Delta t}{\gamma\tau_{\pi_0}}\Big( \delta\pi_i^{n+\frac{1}{2}}-\delta\pi_{NS,i}^{n+\frac{1}{2}}-\xi_i^n \Big)
    \end{split}
\end{equation}
where $\gamma$ is the Lorentz factor of the background velocity and the Navier-Stokes limit entering the relaxation term is
\begin{equation}
    \delta\pi_{NS,i}^{n}=-\frac{4}{3}\eta_0\frac{\delta u_{i+1}^{n}-\delta u_{i-1}^{n}}{2\Delta x}
\end{equation}
We assume the noise in both predictor and corrector for shear stress tensor.
The full step is then realized by adding the shear stress contribution
\begin{equation}
\begin{split}
    \delta\varepsilon_i^{n+1}&=\delta\varepsilon_i^{n+1*}\\
    \delta g_i^{n+1}&=\delta g_i^{n+1*}+\frac{\Delta t}{2\Delta x}(\delta\pi_{i-1}^{n+\frac{1}{2}}-\delta\pi_{i+1}^{n+\frac{1}{2}})
\end{split}
\end{equation}
Now, we can notice that without the noise both predictor and corrector steps have the structure
\begin{equation}
X_i^{n+\frac{1}{2}}=X_i^{n}+\frac{1}{2}\Delta tL X_i^{n}
\end{equation}
and
\begin{equation}
X_i^{n+1*}=X_i^{n}+\Delta tLX_i^{n+\frac{1}{2}}
\end{equation}
respectively, where $L$ is differential operator. 
The full step will be written
\begin{equation}
X_i^{n+1}=X_i^{n}+\Delta tL^{step}X_i^{n+\frac{1}{2}}+\Delta tL^{IS}X_i^{n+\frac{1}{2}}
\end{equation}
where we specifically denoted differential operators $L^{step}$, responsible for evolution of $\delta\varepsilon$ and $\delta g$, and  $L^{IS}$, which is responsible for the evolution of $\delta \pi$. 
Therefore for the full step
\begin{multline}
    X_i^{n+1}=X_i^n+\Delta tL^{step}\Big(I+\frac{1}{2}\Delta tL^{step}\Big)X_i^{n}+\Delta t L^{IS}\Big(I+\frac{1}{2}\Delta tL^{step}\Big)X_i^n\\
    =\Big[\underbrace{ I+\Delta t(L^{step}+L^{IS})+\frac{1}{2}\Delta t^2\left ( (L^{step})^2+L^{IS}L^{step}\right )}_{M} \Big]X_i^n
\end{multline}
Next, we perform the Fourier transform
\begin{multline}
    X_k^{n+1}=X_k^n+\Delta tL_k^{step}\Big(I+\frac{1}{2}\Delta tL_k^{step}\Big)X_k^{n}+\Delta t L_k^{IS}\Big(I+\frac{1}{2}\Delta tL_k^{step}\Big)X_k^n\\
    =\Big[{ I+\Delta t(L_k^{step}+L_k^{IS})
    +\frac{1}{2}\Delta t^2
    ({L_k^{step}}^2+L_k^{IS}L_k^{step})}\Big]X_k^n
\end{multline}
The expression in square brackets defines the matrix $M_k$
\begin{equation}
    M_k = I+\Delta t(L_k^{step}+L_k^{IS})+\frac{1}{2}\Delta t^2\left ((L_k^{step})^2+L_k^{IS}L_k^{step}\right )\, ,
    \label{e:ab:Mk}
\end{equation}
where the operators are given by
\begin{equation}
L_k^{\mathrm{step}} =
\begin{pmatrix}
 -c_s \frac{2\sin^2\left(\frac{k\Delta x}{2}\right)}{\Delta x} & -i\frac{\sin(k\Delta x)}{\Delta x} & 0 \\
 -i c_s^2 \frac{\sin(k\Delta x)}{\Delta x} & -c_s \frac{2\sin^2\left(\frac{k\Delta x}{2}\right)}{\Delta x} & 0 \\
 0 & -\frac{4}{3}\frac{\eta_0}{w_0}\frac{i\sin(k\Delta x)}{\Delta x}\frac{1}{\tau_{\pi_0}} & -\frac{1}{\tau_{\pi_0}}
\end{pmatrix},
\end{equation}
\begin{equation}
L_k^{\mathrm{IS}} =
\begin{pmatrix}
 0 & 0 & 0 \\
 0 & 0 & -i\frac{\sin(k\Delta x)}{\Delta x} \\
 0 & 0 & 0
\end{pmatrix}.
\end{equation}
\end{widetext}
The stochastic noise enters only through the shear stress evolution. The corresponding noise matrix is
\begin{equation}
    N_k =
    \begin{pmatrix}
        0 & 0 & 0 \\
        0 & 0 & 0 \\
        0 & 0 &
        \frac{\Delta t}{\tau_{\pi_0}}
        \left(1 - \frac{\Delta t}{2\tau_{\pi_0}}\right)\sigma
    \end{pmatrix},
\end{equation}
where
\begin{equation}
    \sigma^2 = \frac{8}{3}\eta_0 T_0 \frac{1}{\Delta x \Delta t}.
\end{equation}
This yields
\begin{equation}
    Q_k =
    \begin{pmatrix}
        0 & 0 & 0 \\
        0 & 0 & 0 \\
        0 & 0 &
        -\frac{8}{3}\eta_0 T_0
        \frac{\Delta t}{\tau_{\pi_0}^2}
        \left(1 - \frac{\Delta t}{2\tau_{\pi_0}}\right)^2
    \end{pmatrix}.
    \label{e:ab:Qk}
\end{equation}
We thus have all the matrices that are put into  the discrete Lyapunov equation
\begin{equation}
    M_k S M_k^\dagger - S = Q_k,
\label{e:e4numSF}
\end{equation}
which can be solved for $S$ numerically via vectorization
\begin{equation}
    (M_k \otimes M_k^\ast - I)\mathrm{vec}(S) = \mathrm{vec}(Q_k).
\end{equation}


\end{document}